\documentstyle[12pt,epsfig]{article}

\catcode`\@=11
\def\@citex[#1]#2{%
\if@filesw \immediate \write \@auxout {\string \citation {#2}}\fi
\@tempcntb\m@ne \let\@h@ld\relax \def\@citea{}%
\@cite{%
  \@for \@citeb:=#2\do {%
    \@ifundefined {b@\@citeb}%
      {\@h@ld\@citea\@tempcntb\m@ne{\bf ?}%
      \@warning {Citation `\@citeb ' on page \thepage \space undefined}}%
      {\@tempcnta\@tempcntb \advance\@tempcnta\@ne%
      \@tempcntb\number\csname b@\@citeb \endcsname \relax%
      \ifnum\@tempcnta=\@tempcntb %Number follows previous--hold on to it
        \ifx\@h@ld\relax%
          \edef \@h@ld{\@citea\csname b@\@citeb\endcsname}%
        \else%
          \edef\@h@ld{\ifmmode{-}\else--\fi\csname b@\@citeb\endcsname}%
        \fi%
      \else%   %  non-successor--dump what's held and do this one
        \@h@ld\@citea\csname b@\@citeb \endcsname%
        \let\@h@ld\relax%
      \fi}%
    \def\@citea{,\penalty\@highpenalty\,}%
  }\@h@ld
}{#1}}

\def\@citeb#1#2{{[#1]\if@tempswa , #2\fi}}
\def\@citeu#1#2{{$^{#1}$\if@tempswa , #2\fi }}
\def\@citep#1#2{{#1\if@tempswa , #2\fi}}

\def\bcites{         % cite with []'s
        \catcode`\@=11
        \let\@cite=\@citeb
        \catcode`\@=12
}

\def\upcites{         % cite with exponents
        \catcode`\@=11
        \let\@cite=\@citeu
        \catcode`\@=12
}

\def\plaincites{      % cite without brackets
        \catcode`\@=11
        \let\@cite=\@citep
        \catcode`\@=12
}

\newcount\hour
\newcount\minute
\newtoks\amorpm
\hour=\time\divide\hour by 60
\minute=\time{\multiply\hour by 60 \global\advance\minute by-\hour}
\edef\standardtime{{\ifnum\hour<12 \global\amorpm={am}%
        \else\global\amorpm={pm}\advance\hour by-12 \fi
        \ifnum\hour=0 \hour=12 \fi
        \number\hour:\ifnum\minute<10 0\fi\number\minute\the\amorpm}}
\edef\militarytime{\number\hour:\ifnum\minute<10 0\fi\number\minute}

\def\draftlabel#1{{\@bsphack\if@filesw {\let\thepage\relax
   \xdef\@gtempa{\write\@auxout{\string
      \newlabel{#1}{{\@currentlabel}{\thepage}}}}}\@gtempa
   \if@nobreak \ifvmode\nobreak\fi\fi\fi\@esphack}
        \gdef\@eqnlabel{#1}}
\def\@eqnlabel{}
\def\@vacuum{}
\def\marginnote#1{}
\def\draftmarginnote#1{\marginpar{\raggedright\scriptsize\tt#1}}
\def\draft{
        \pagestyle{plain}
        \overfullrule=2pt
        \oddsidemargin -.5truein
        \def\@oddhead{\sl \phantom{\today\quad\militarytime} \hfil
        \smash{\Large\sl DRAFT} \hfil \today\quad\militarytime}
        \let\@evenhead\@oddhead
        \let\label=\draftlabel
        \let\marginnote=\draftmarginnote
        \def\ps@empty{\let\@mkboth\@gobbletwo
        \def\@oddfoot{\hfil \smash{\Large\sl DRAFT} \hfil}
        \let\@evenfoot\@oddhead}
        \def\@eqnnum{(\theequation)\rlap{\kern\marginparsep\tt\@eqnlabel}%
        \global\let\@eqnlabel\@vacuum}  }

\def\blackfonts{
        \font\blackboard=msbm10 scaled\magstep1
        \font\blackboards=msbm8
        \font\blackboardss=msbm6
}

\def\nblack{            % For people without blackboard fonts
        \def\ZZ{{Z \n{10} Z}}
        \def\NN{{N \n{14} N}}
        \def\CC{{C \n{11} C}}
        \def\RR{{R \n{11} R}}
        \def\QQ{{Q \n{12} Q}}
        \def\PP{{P \n{11} P}}
}

\def\prep{         % twocolumn.sty  Changed by Marek and Neil
        \catcode`\@=11
        \input art10.sty
        \catcode`\@=12
        
        \let\small\null
        \def\blackfonts{
                \font\blackboard=msbm10
                \font\blackboards=msbm7
                \font\blackboardss=msbm5
        }
        \let\sl\it
        \twocolumn
        \sloppy
        \voffset=-2.54truecm
        \hoffset=-2.54truecm
        \flushbottom
        \parindent 1em
        \leftmargini 2em
        \leftmarginv .5em
        \leftmarginvi .5em
        \marginparwidth 48pt
        \marginparsep 10pt
        \setlength{\columnsep}{2truecm}
        \setlength{\textwidth}{25.4truecm}
        \setlength{\textheight}{17truecm}
        \baselineskip=16pt
        \oddsidemargin .18truein
        \evensidemargin .17truein
}

\def\eqalign#1{\null\,\vcenter{\openup\jot\m@th
  \ialign{\strut\hfil$\displaystyle{##}$&$\displaystyle{{}##}$\hfil
      \crcr#1\crcr}}\,}
\def\eqalignno#1{\displ@y \tabskip\centering
  \halign to\displaywidth{\hfil$\@lign\displaystyle{##}$\tabskip\z@skip
    &$\@lign\displaystyle{{}##}$\hfil\tabskip\centering
    &\llap{$\@lign##$}\tabskip\z@skip\crcr
    #1\crcr}}

\def\section{\@startsection {section}{1}{\z@}{3.ex plus 1ex minus
 .2ex}{2.ex plus .2ex}{\large\bf}}
\def\subsection{\@startsection{subsection}{2}{\z@}{2.75ex plus 1ex minus
 .2ex}{1.5ex plus .2ex}{\bf}}        

\def\appendix{{\newpage\section*{Appendix}}\let\appendix\section%
        {\setcounter{section}{0}
        \gdef\thesection{\Alph{section}}}\section}

\def\abstract{\if@twocolumn
\section*{Abstract}
\else %\small
\begin{center}
{\bf Abstract\vspace{-.5em}\vspace{0pt}}
\end{center}
\quotation
\fi}

\catcode`\@=12

\newcommand{\beq}{\begin{equation}}
\newcommand{\eeq}{\end{equation}}
\newcommand{\beqa}{\begin{eqnarray}}
\newcommand{\eeqa}{\end{eqnarray}}

\newcommand{\Z}{{\bf Z}}

\newcommand{\R}{{\bf R}}
\newcommand{\C}{{\bf C}}
\newcommand{\e}{{\rm e}}

\newcommand{\dd}{{\rm d}}
\newcommand{\elst}{{\ell_{\it st}}}
\newcommand{\elel}{{\ell_{11}}}
\newcommand{\gst}{{g_{\it st}}}
\newcommand{\MT}{{$M$ theory}~}

\def\noj#1,#2,{{\bf #1} (19#2)\ }
\def\jou#1,#2,#3,{{\sl #1\/ }{\bf #2} (19#3)\ }
\def\ann#1,#2,{{\sl Ann.\ Physics\/ }{\bf #1} (19#2)\ }
\def\cmp#1,#2,{{\sl Comm.\ Math.\ Phys.\/ }{\bf #1} (19#2)\ }
\def\ma#1,#2,{{\sl Math.\ Ann.\/ }{\bf #1} (19#2)\ }
\def\ng#1,#2,{{\sl Nagoya.\ Math.\ J.\/ }{\bf #1} (19#2)\ }
\def\jd#1,#2,{{\sl J.\ Diff.\ Geom.\/ }{\bf #1} (19#2)\ }
\def\invm#1,#2,{{\sl Invent.\ Math.\/ }{\bf #1} (19#2)\ }
\def\cq#1,#2,{{\sl Class.\ Quantum Grav.\/ }{\bf #1} (19#2)\ }
\def\cqg#1,#2,{{\sl Class.\ Quantum Grav.\/ }{\bf #1} (19#2)\ }
\def\ijmp#1,#2,{{\sl Int.\ J.\ Mod.\ Phys.\/ }{\bf A#1} (19#2)\ }
\def\jmphy#1,#2,{{\sl J.\ Geom.\ Phys.\/ }{\bf #1} (19#2)\ }
\def\jams#1,#2,{{\sl J.\ Amer.\ Math.\ Soc.\/ }{\bf #1} (19#2)\ }
\def\grg#1,#2,{{\sl Gen.\ Rel.\ Grav.\/ }{\bf #1} (19#2)\ }
\def\mpl#1,#2,{{\sl Mod.\ Phys.\ Lett.\/ }{\bf A#1} (19#2)\ }
\def\nc#1,#2,{{\sl Nuovo Cim.\/ }{\bf #1} (19#2)\ }
\def\np#1,#2,{{\sl Nucl.\ Phys.\/ }{\bf B#1} (19#2)\ }
\def\pl#1,#2,{{\sl Phys.\ Lett.\/ }{\bf #1B} (19#2)\ }
\def\pla#1,#2,{{\sl Phys.\ Lett.\/ }{\bf #1A} (19#2)\ }
\def\pr#1,#2,{{\sl Phys.\ Rev.\/ }{\bf #1} (19#2)\ }
\def\prd#1,#2,{{\sl Phys.\ Rev.\/ }{\bf D#1} (19#2)\ }
\def\prl#1,#2,{{\sl Phys.\ Rev.\ Lett.\/ }{\bf #1} (19#2)\ }
\def\prp#1,#2,{{\sl Phys.\ Rept.\/ }{\bf #1C} (19#2)\ }
\def\ptp#1,#2,{{\sl Prog.\ Theor.\ Phys.\/ }{\bf #1} (19#2)\ }
\def\ptpsup#1,#2,{{\sl Prog.\ Theor.\ Phys.\/ Suppl.\/ }{\bf #1} (19#2)\ }
\def\rmp#1,#2,{{\sl Rev.\ Mod.\ Phys.\/ }{\bf #1} (19#2)\ }
\def\yadfiz#1,#2,#3[#4,#5]{{\sl Yad.\ Fiz.\/ }{\bf #1} (19#2) #3%
\ [{\sl Sov.\ J.\ Nucl.\ Phys.\/ }{\bf #4} (19#2) #5]}
\def\zh#1,#2,#3[#4,#5]{{\sl Zh.\ Exp.\ Theor.\ Fiz.\/ }{\bf #1} (19#2) #3%
\ [{\sl Sov.\ Phys.\ JETP\/ }{\bf #4} (19#2) #5]}

\def\beq{\begin{equation}}
\def\eeq{\end{equation}}
\def\beqar{\begin{eqnarray}}
\def\eeqar{\end{eqnarray}}

\newcommand{\be}{\begin{equation}}
\newcommand{\ee}{\end{equation}}
\newcommand{\bea}{\begin{eqnarray}}
\newcommand{\eea}{\end{eqnarray}}

\def\nfrac#1#2{{\displaystyle{\vphantom1\smash{\lower.5ex\hbox{\small$#1$}}%
        \over\vphantom1\smash{\raise.25ex\hbox{\small$#2$}}}}}

\def\n#1{\mskip-#1mu}

\def\to{\rightarrow}

\def\lae{\mathrel{\mathop{\smash{\lower .5 ex \hbox{$\stackrel<\sim$}}}}}
\def\lae{\mathrel{\mathop{\smash{\lower .5 ex \hbox{$\stackrel>\sim$}}}}}

\def\l:{\mathopen{:}\,}
\def\r:{\,\mathclose{:}}

\catcode`\@=11
\def\theequation{\arabic{equation}}
\catcode`\@=12

\nblack
\bcites

\nblack

\catcode`\@=11
\def\theequation{\thesection.\arabic{equation}}
\@addtoreset{equation}{section}
\@addtoreset{footnote}{section}
\@addtoreset{footnote}{subsection}
\catcode`\@=12

\newcommand{\beqn}{\begin{equation}}
\newcommand{\eeqn}{\end{equation}}
\newcommand{\beqnarray}{\begin{eqnarray}}
\newcommand{\eeqnarray}{\end{eqnarray}}
\newcommand {\bear} [1] {\begin {array} {#1}}
\newcommand {\ear} {\end {array}}

\newcommand{\CP}{{\bf C}{\rm P}}
\newcommand{\RP}{{\bf R}{\rm P}}

\newcommand {\beqarn} {\begin{eqnarray*}}
\newcommand {\eeqarn} {\end{eqnarray*}}

\begin{document}

\begin{titlepage}

\begin{center}
\today
\hfill LBNL-41736, UCB-PTH-98/22\\
\hfill                  hep-th/9805141

\vskip 1.5 cm
{\large \bf Consistency Conditions for Fivebrane
 in $M$ Theory\\[0.2cm]
on $\R^5/\Z_2$ Orbifold}
\vskip 1 cm 
{Kentaro Hori}\\
\vskip 0.5cm
{\sl Department of Physics,
University of California at Berkeley\\
366 Le\thinspace Conte Hall, Berkeley, CA 94720-7300, U.S.A.\\
and\\
Theoretical Physics Group, Mail Stop 50A--5101\\
Ernest Orlando Lawrence Berkeley National Laboratory\\
Berkeley, CA 94720, U.S.A.\\}

\end{center}

\vskip 0.5 cm
\begin{abstract}
\noindent
We derive some consistency conditions for fivebrane in \MT
on $\R^5/\Z_2$ orbifold from the quantization law
for the antisymmetric tensor field.
We construct consistent fivebrane configurations in $\R^5/\Z_2$ type
orbifold that exhibit the correct low energy
dynamics of $N=2$ SQCD in four dimensions with symplectic and orthogonal
gauge groups.
This leads us to propose the \MT realization of orientifold four-planes
of various types, and we study their properties by applying the
consistency conditions.

\end{abstract}

\end{titlepage}

\section{Introduction}

The Dirac quatization of
electric and magnetic charges is a strict and
powerful condition for the consistency of a theory.
In $M$ theory, membrane and fivebrane are the electrically
and magnetically charged objects with respect to the three-form
gauge potential $C$.
The Dirac quantization condition for these objects
takes the form of flux quantization condition \cite{Wflux}
for the four-form field strength $G$:
\beq
2\int\limits_S{G\over 2\pi}~\equiv ~ \int\limits_S{w_4}
~~~~~\mbox{mod 2},
\label{Flux}
\eeq
where $S$ is a four-cycle and
$w_4$ is the fourth Stiefel Whitney class of the eleven-dimensional
space-time.
This has a direct consequence in \MT on $\R^5/\Z_2$ orbifold
\cite{Wflux}.
For a four-cycle surrounding the origin of $\R^5/\Z_2$
the right hand side
of (\ref{Flux}) is one mod 2, and therefore, the flux of
$G/2\pi$ through the four-sphere surrounding the $\Z_2$ fixed plane
in the double cover must always be odd.
This is consistent with the result \cite{DM,orbW} that the
$\Z_2$ fixed plane itself carries the fivebrane charge
$-1$.
Moreover, the condition implies that it is inconsistent to have
odd number of fivebranes on top of the $\Z_2$ fixed plane.
In this paper,
we derive from the quantization condition (\ref{Flux})
more general consistency
conditions for configuration of the
fivebrane in \MT on orbifold of $\R^5/\Z_2$ type.

Lately, low energy dynamics of supersymmetric gauge theories in
various dimensions have been studied by realizing them on the
worldvolume of branes and using string duality (see \cite{GK}
and references therein).
In particular, theories in four-dimensions can be realized on certain
configurations of branes in Type IIA string theory and some of the
features that are independent of the string coupling constant
can be effectively studied by lifting them to configurations of
\MT fivebrane.
When the characteristic length of the fivrbrane and the space-time is
larger than the eleven-dimensional Planck length, the low energy
supergravity approximation of \MT can be used reliably.
In some important cases, however, the configuration is in a space-time
that has orbifold singularity of $\R^5/\Z_2$ type, and the description
based on the supergravity approximation breaks down when the fivebrane
becomes close to the singular points.
This is in particular the case when the gauge symmetry in four
dimensions is the symplectic or orthogonal group which is realized by
introducing orientifold four-plane in Type IIA string
theory.
\footnote{An alternative way to realize these groups is to use
orientifold six-plane. The one that leads to orthogonal flavor group
can be realized as \MT on a smooth geometry \cite{SeibergIR,SW}, and
therefore it can be used reliably to study symplectic gauge theories
\cite{HOV}. Attempts to study orthogonal gauge theories in a similar
way suffer from a mysterious inherent singularity (see for example
\cite{LL,Jaemo}).}
This work
is motivated by the need to have a control over the
behaviour of the fivebrane in such a singular background.

The orientifold four-planes of various types,
especially their \MT realization and the behaviour upon
intersection with other branes, have not been well understood.
Therefore, we did not actually know the precise Type IIA configuration
that realize the symplectic or
orthogonal gauge theory,\footnote{See
\cite{EJS,EGKRS,GK} for some proposals.
(We will find that these are incorrect in the detail.)}
let alone the lift to $M$ theory.
In such a situation, reversing the direction of the logic,
we may first find a fivebrane configuration in \MT that have the
correct low energy properties of the gauge theory and then take the
weak string coupling limit to know about the orientifold four-plane
and the relevant Type IIA configuration.
This program did not make sense
in the past, since we had no control over the fivebrane
in a singular background.
However, once we know a criterion for
consistent configurations,
this program will eventually lead to the \MT realization
of orientifold four-planes and to an understanding of their
properties. This will be the goal of this paper.

There were several proposals \cite{LLL,Tera,DSB}
for fivebrane configurations in $\R^5/\Z_2$ orbifold
that correspond to
theories with symplectic or orthogonal groups,
but extra assumptions about the behaviour of branes near the
$\Z_2$ fixed plane were required in order to have a consistent result
with the ordinary field theory.
In \cite{DSB}, it was noticed that the basic assumption that has to be
made (at least in the case of symplectic gauge group)
is that a certain intersection of the fivebrane and the
$\Z_2$ fixed plane --- which we call {\it t}-configuration ---
breaks supersymmetry:
Only from this, one can derive most of the essential features of
theories with symplectic gauge group, such as
the Seiberg-Witten curve
of $N=2$ theories \cite{ASh,DKPh} (as was first observed in \cite{LLL}),
quantum modified constraint in $N=1$ SQCD \cite{Exact,IP},
and dynamical supersymmetry breaking of the models of \cite{IY,IT}.
In section 2, we examine a {\it t}-configuration
and show that it
does not satisfy the flux quantization condition (\ref{Flux}).
Namely, we show that a {\it t}-configuration is
not only non-supersymmetric but is actually inconsistent.
We also derive more general consistency conditions.

In section 3, we see the implication of the consistency conditions
in string theory, by considering a {\it t}-configuration and its
cousins in a toroidally compactified space-time.
We will see that some of them are related to known or
conjectured consistency conditions in string theory.

In \cite{DSB}, fivebrane configuration
in the presence of sixbranes was studied, and
an additional assumption generalizing the
basic assumption mentioned above was proposed
in order to have the correct dimension of the Higgs branch of
symplectic gauge theory.
In section 4, we point out that the brane configuration constructed
in \cite{DSB} has an inconsostent part, 
and show that there is no need to assume anything once we modify the
brane configuration appropriately.
The flux quantization condition is sufficient to
select consistent configurations and we obtain results
in agreement with field theory.

In section 5 we construct and analyze
the brane configuration corresponding
to $N=2$ theories with orthogonal gauge groups.
Using the consistency conditions derived in section 2,
we show that the configuration we will find reproduces the results
of field theory.
In the course of construction, we will discover new
\MT realization of orientifold four-planes.
We also encounter a phenomenon where fivebranes are created when two
coincident sixbranes are separated while intersecting with the
orientifold four-plane.

In section 6, the \MT realization of orientifold four-planes
is examined in more detail. We show that orientifold four-plane of
$SO({\rm even})$-type has trivial RR gauge field
whereas $SO({\rm odd})$-type has non-trivial RR holonomy.
We also show that there are two
orientifold four-planes of $Sp$-type --- one with trivial RR $U(1)$
gauge field and one with non-trivial holonomy.
We discuss the possibility of a global term
in the CS coupling in the D-brane effective action
(coupling of the RR potentials and gauge field on the D-brane)
such that the two constructions of $Sp$-type O4-plane
correspond to the two choices of $\Z_2$ valued theta angle
of five-dimensional symplectic gauge theory in $4+1$ dimensions.

In section 7, as an application of the consistency conditions
and the \MT realization of orientifold four-planes, we
study the properties of orientifold four-planes intersecting with
D6-branes and/or NS5-branes. In particular, we derive the
brane creation rule and the $s$-rule in the presence of orientifold
four-plane.

{\it In this paper, we usually count the number of fivebranes
of \MT on $\Z_2$ orbifold (or branes in Type II orientifold)
in the double cover of the $\Z_2$ quotient,
unless otherwise stated.}

\section{Fivebrane in $\R^5/\Z_2$ Orbifold}

\newcommand{\hX}{\widehat{X}}
\newcommand{\hS}{\widehat{S}}
\newcommand{\tilR}{\widetilde{\R}}
\newcommand{\Zo}{\Z^{\mbox{\tiny $\cal O$}}}

In this section, we derive the consistency condition for the fivebrane
in $\R^5/\Z_2$ orbifold using the flux quantization condition
(\ref{Flux}).

We consider the eleven-dimensional space-time $X$ which
is the $\Z_2$ quotient of an orientable manifold $\hX$,
where the action of the generator $\gamma$ of $\Z_2$
is orientation reversing and has the $\R^5/\Z_2$ type fixed points.
Namely, $\hX$ has a six-dimensional submanifold of $\Z_2$ fixed points
(which we shall call ``$\Z_2$-fixed plane'' even when it is curved)
along which $\gamma$ acts as the sign flip of the five transverse
coordinates.
We shall usually coordinatize the $\Z_2$-fixed plane by
$x^{0,1,2,3,6,10}$ and denote the transverse coordinates by
$x^{4,5,7,8,9}$.
$X$ must be a pin-manifold in order for \MT to be formulated on it,
and therefore its double cover $\hX$ must have a spin-structure.

Since the $\Z_2$ action flips the orientation of $\hX$,
the field strength $G$ on $\hX$ flips its sign under the $\Z_2$ action,
and the orbifolding is possible only when $\gamma^*G=-G$.
Then, $G$ does not define an ordinary four-form in the quotient
space $X=\hX/\Z_2$, but rather,
a four-form with values in the orientation bundle $o_X$ of $X$,
the real determinant line bundle of the tangent bundle of
$X$. ($o_X$ has a transition function with values in $\pm 1$ and therefore
is reducible to a $\Z$-bundle).
Even in such a case,
the flux of $G$ through a submanifold
can be defined as a real number when the double cover of the
submanifold is oriented in a particular way.
Let $\hS$ be a $\Z_2$-invariant four-dimensional submanifold of $\hX$
with an orientation that is flipped under the $\Z_2$ action.
Then, the flux of $G/2\pi$ through its $\Z_2$ quotient,
$S\subset X$, is defined as half of the flux through $\hS$,
$\int_S{G\over 2\pi}:={1\over 2}\int_{\hS}{G\over 2\pi}$.
Of course, the integral of $w_4(X)$ over $S$
is defined only as a mod 2 integer (an element of $\Z/2\Z$),
and the condition (\ref{Flux}) should be understood as saying
that $2\int_SG/2\pi$ is an integer and its
mod 2 parity agrees with $\int_Sw_4(X)$.\footnote{
More formally,
the condition (\ref{Flux}) first of all implies
that $G/\pi$ belongs to a fourth cohomology class $[G/\pi]$
with values in the twisted $integer$ coefficient $\Zo$
(namely an element in $H^4(X,\Zo)$) where the twisting
is determined by the orientation bundle $o_X$.
For such a class, the integral over an arbitrary not necessarily
oriented submanifold of $X$ can be defined as a mod 2 integer.
The condition (\ref{Flux}) states that it is the same as the integral
of $w_4(X)$.
In other words, the mod 2 reduction
of $[G/\pi]\in H^4(X,\Zo)$ is equal to $w_4(X)\in H^4(X,\Z/2\Z)$.}

For instance, let us consider \MT on $\R^5\times \R^5/\Z_2$.
Suppose there are a pair of fivebranes in $\hX=\R^{10}$ parallel to
but separate from the $\Z_2$-fixed plane which are mirror images
of each other.
We pick one four-sphere $S^4_{(1)}$ surrounding one
of them together with its mirror image $S^4_{(2)}$,
and give each of them the orientation
that is induced from the orientation of $\R^5$.
Put $\hS=S^4_{(1)}\cup S^4_{(2)}$.
Then, the orientation of $\hS$ is fliped by $\Z_2$
since the orientation of $S^4_{(1)}$ and $S^4_{(2)}$
is chose so that $\gamma_*[S^4_{(1)}]=-[S^4_{(2)}]$.
Then, the flux through the quotient $S$
is given by
$\int_SG/2\pi={1\over 2}(\int_{S^4_{(1)}}{G\over 2\pi}+
\int_{S^4_{(2)}}{G\over 2\pi})={1+1\over 2}=1$,
as in the usual case where there is a single fivebrane 
in an orientable space-time.
On the other hand, the tangent bundle of the space-time
is topologically trivial around the four-sphere surrounding the
fivebrane, and thus $\int_Sw_4=0$ mod 2. Therefore,
the flux quantization condition (\ref{Flux}) is satisfied.

This argument applies to the case of general $X$ and
we can show
that there is no obstruction from the flux quantization condition
to having a pair of fivebranes as far as they are separated from the
$\Z_2$ fixed plane.

\subsection{Fivebranes on Top of the $\Z_2$ Fixed Plane}

This subsection is a review of section 2.3 of the paper \cite{Wflux}.
We consider \MT on $\R^6\times \R^5/\Z_2$.
Since the supergravity approximation may not be valid near the
singularity at the $\Z_2$ fixed plane, we delete a neighborhood of
the origin of $\R^5/\Z_2$
and denote the resulting space by $\tilR^5/\Z_2$.

We take a four-cycle $S$ in $\tilR^5/\Z_2$
surrounding the (deleted) $\Z_2$ fixed plane.
To be specific, we take a large four-sphere $S^4$ surrounding the
origin of $\R^5$ and give it a natural orientation.
$S^4$ is $\Z_2$ invariant and its orientation is reversed under
$\gamma$.
Thus, the flux of $G/2\pi$ through the quotient $S=S^4/\Z_2$
is defined as half of the flux through $S^4$.
The quotient $S^4/\Z_2$ is isomorphic to
the real projective space $\RP^4$
and the mod 2 homology class of $S$ is the image of the mod 2
fundamental class of $\RP^4$.
The tangent bundle of $\tilR^5/\Z_2$ restricted to $S$
is the direct sum of five copies of an unorientable
real line bundle $m$ whose first Stielfel Whitney
class generates the cohomology ring over $\Z_2$ of $\RP^4$.
Since the total Stiefel Whitney class is
$(1+w_1(m))^5=1+w_1(m)+w_1(m)^4$ (see the forthcoming footnote),
the integral of $w_4$ on $S$ is equal to one mod 2.
Therefore the flux quantization condition (\ref{Flux})
requires
\beq
2\int\limits_{S}{G\over 2\pi}\,\equiv\,1~~~{\rm mod}~~2.
\label{fll}
\eeq
This means that the flux of $G/2\pi$ through a four-sphere
surrounding the $\Z_2$ fixed plane is odd when measured in
the double cover.

This is true even when the $\Z_2$ fixed plane
is curved. We pick an arbitrary point in the curved $\Z_2$ fixed plane
and take a normal five-plane passing through it on which the $\Z_2$ acts
as the sign flip. Then, we can take
a $\Z_2$ invariant four-sphere $S^4$ in the normal five-plane.
The result (\ref{fll}) is the same since the integral of
$w_4$ on $S^4/\Z_2$
is purely determined by the restriction on $S^4/\Z_2$
of the tangent bundle of the eleven-dimensional space-time
which is isomorphic to the bundle over $\RP^4$
considered above.

In \cite{DM,orbW}, \MT compactified on $T^5/\Z_2$ was studied
and it was concluded from local cancellation of gravitational anomaly
that each $\Z_2$ fixed plane
has the fivebrane charge $-1$
(when counted before the $\Z_2$ quotient). 
Namely, for an $S^4$ in $T^5$ surrounding the $\Z_2$ fixed
plane, we have $\int_{S^4}G/2\pi=-1$.\footnote{
This can also be seen via the quantum mechanics
of zero-brane probes of orientifold four-plane \cite{Rey}.}
This is consistent with
the flux quantization condition considered above.
Furthermore, one cannot locate an odd number of fivebranes
(from the total of 32) at any of $\Z_2$ fixed plane
since cancellation of six-dimensional gravitational anomaly
requires 16 tensor multiplets which can only be provided
by freely moving 16 pairs of fivebranes.
This is also consistent with the flux quantization condition;
if an odd number of fivebranes were on top of the
$\Z_2$ fixed plane, the flux through a four-sphere surrounding the
fixed plane would be even, in contradiction with (\ref{fll}).

{\bf To summarize}, {\it
the $\Z_2$ fixed plane carries fivebrane charge $-1$
and one cannot put odd number of fivebranes on top of it.}

\subsection{Fivebranes Intersecting with the $\Z_2$ Fixed Plane}

Now we consider fivebranes intersecting transversely with
the $\Z_2$ fixed plane.
Let us call a ``{\it t}-configuration''
an intersection of a single fivebrane and the $\Z_2$ fixed plane of the
following type:
the $\Z_2$ fixed plane is at
$x^{4,5,7,8,9}=0$ and spans the $x^{0,1,2,3,6,10}$-directions,
while the fivebrane is at $x^{6,7,8,9,10}=0$
and spans the worldvolume in the $x^{0,1,2,3,4,5}$ directions.
They both lie in $x^{7,8,9}=0$
and span the common directions $x^{0,1,2,3}$.
In the remaining directions $x^{4,5,6,10}$,
the fivebrane spanning $x^{4,5}$ and
the $\Z_2$ fixed plane spanning
$x^{6,10}$ intersect transversely at $x^{4,5,6,10}=0$.
Apparently this configuration preserves eight
supersymmetry generators because there are eight constant spinors with
$\Gamma^0\Gamma^1\Gamma^2\Gamma^3\Gamma^4\Gamma^5
=\Gamma^0\Gamma^1\Gamma^2\Gamma^3\Gamma^6\Gamma^{10}=1$
where $\Gamma^{i}$ are the eleven-dimensional gamma matrices.
In \cite{DSB}, it was conjectured that a {\it t}-configuration
actually breaks all of the supersymmetries.
In this subsection we show that it is not only non-supersymmetric but
actually is inconsistent because it does not satisfy the flux
quantization condition of $M$ theory (\ref{Flux}).
We also find a more general consistency condition.

Let us temporarily consider the case where the $x^6$ and $x^{10}$
directions are compactified on a torus $T^2$, $x^6\equiv x^6+2\pi R_1$
$x^{10}\equiv x^{10}+2\pi R_2$.
We take the four-cycle $\hS=T^2\times S^2$ in $x^{4,5}=0$
where $T^2$ is the torus in the $x^{6,10}$ directions
and $S^2$ is a two-sphere $|x^{7,8,9}|=R$
in the three-plane $\R^3$ spanning the $x^{7,8,9}$-directions
(we will work always at a single point in the $x^{0,1,2,3}$ directions
and this will not be mentioned in what follows).
The $\Z_2$ action reverses the orientation of $\R^3$
and hence the orientation of the cycle $\hS$
is also flipped.
Thus, the flux of $G/2\pi$ through the $\Z_2$ quotient
$S=T^2\times\RP^2$ is defined as half of the flux though
$\hS=T^2\times S^2$.

Let us measure the flux of $G/2\pi$ over this $T^2\times S^2$.
For this purpose
one can translate the cycle from $x^{4,5}=0$ to somewhere with 
$x^{4,5}\ne 0$.
The flux does not change since the cycle does not pass
through the locus of the fivebrane $x^{6,7,8,9,10}=0$
nor the $\Z_2$ fixed plane $x^{4,5,7,8,9}=0$ because $|x^{7,8,9}|$
is always non-zero in the process of translation.
Then, let us consider a circle $|x^{6,7}|=\epsilon$ in $T^2$ surrounding
$x^{6,10}=0$.
We deform the cycle $T^2\times S^2$
so that the two-sphere $S^2$ is pinched along the circle
(here we consider $T^2\times S^2$ as $S^2$-bundle over $T^2$).
Namely, we change the equation
$|x^{7,8,9}|=R$ to $|x^{7,8,9}|=Rf(x^{6,10})$ where
$f(x^{6,10})$ is a function which is zero at $|x^{6,10}|=\epsilon$
but is positive everywhere else.
By this deformation, obviously the flux is not changed.
Then, the cycle $T^2\times S^2$ splits to two components;
one is inside the circle $|x^{6,7}|\leq\epsilon$
and is a four-sphere surrounding
the fivebrane at $x^{6,7,8,9,10}=0$,
and the other is outside the circle $|x^{6,7}|\geq\epsilon$
and surrounds nothing ---
neither fivebrane nor the $\Z_2$ fixed plane.
The flux through the former component is of course one
and the flux through the latter is obviously zero.
Thus, the flux of $G/2\pi$
through $T^2\times S^2$
is equal to one, or equivalently
\beq
2\int\limits_{T^2\times \RP^2}{G\over 2\pi}~=~1\,.
\eeq
Therefore, the flux quantization condition (\ref{Flux}) requires
that the integral of $w_4$ over $T^2\times \RP^2$ is half-integer.

On the other hand, one can show that
\beq
\int\limits_{T^2\times \RP^2}w_4=0~~{\rm mod}~2\,,
\eeq
as we will see in a more general setting shortly.
This contradicts with the requirement.
Thus, the flux quantization condition (\ref{Flux})
does not hold for $S=T^2\times \RP^2$.

Therefore we conclude that a {\it t}-configuration is inconsistent
when the $\Z_2$ fixed plane is compactified on a torus $T^2$.
By taking the infinite volume limit $R_1,R_2\to\infty$
one can equally say that a {\it t}-configuration is inconsistent
for an infinite $\Z_2$ fixed plane
when there is no charge flow from infinity.
Likewise, it is not possible to have
the fivebrane of odd charge intersecting with the
$\Z_2$ fixed plane.

If the charge of the fivebranes is even,
the condition (\ref{Flux}) is satisfied for
$S=T^2\times\RP^2$.
Actually, such a configuration arizes as a limit of
a consistent configuration.
Let us consider two sheets of the fivebrane at $x^{6,7,8,9,10}=0$,
remove the parts with $|x^{4,5}|< \epsilon$
and glue them along
the boundaries at $|x^{4,5}|=\epsilon$.
As mentioned at the begining of this section, there is no
inconsistency
when the fivebrane is away from the $\Z_2$ fixed plane.
If $\epsilon$ is taken to be as small as
the eleven-dimensional Planck length $\elel$,
the configuration looks like two fivebranes on top of each other
intersecting with the $\Z_2$ fixed plane.
The two sheets can be rotated in
the $x^{4,5,7,8,9}$-directions in different ways.
For example, one sheet can be rotated to $x^{4,5,9}=0$
while the other remains in $x^{7,8,9}=0$.
This kind of configuration
can also be realized as a limit of a consistent
configuration.\footnote{
If we introduce complex coordinates $v=x^4+ix^5$
and $w=x^8+ix^9$, the intersecting fivebranes can be described by
$vw=0$, $x^{6,7,10}=0$. This arizes as a limit of the configuration
$vw=\epsilon$ where
the fivebrane is away from the $\Z_2$ fixed plane.}
Therefore, we conclude that it is possible to have
even number of fivebranes (in arbitrary directions)
intersecting transversely with the $\Z_2$ fixed plane
at the same point.

\subsection*{\sl $\R^5/\Z_2$ Orbifold along a Riemann Surface}

We consider here a more general case where the $\Z_2$ fixed plane is
wrapped on a curved two-dimensional surface.
Namely, we consider \MT on $\R^4\times M^7/\Z_2$ where $M^7$
is a seven-dimensional spin manifold and the $\Z_2$ action on it
has fixed points along a compact orientable
two-dimensional submanifold $\Sigma\subset M^7$.
In later sections where we will use the result
of the present section, we will encounter the case where
$M^7=K\times \R^3$ where $K$ is a multi Taub-NUT space and the $\Z_2$-fixed
plane is the union of some rational curves in $K$ at the origin of $\R^3$.
As in such a case, $\Sigma$ can in general have several components.
However, we will only focus on a neighborhood of a single
component in the present discussion, so we may as well assume
that $\Sigma$ is a single smooth Riemann surface.
Furthermore, we can approximate $M^7$ by the total space of
the normal bundle $N$ of $\Sigma$ in $M^7$ on which the
$\Z_2$ acts as the sing flip on the fibre.
The rank of $N$ is of course five.

Let us consider a configuration where $n$ fivebranes intersect
transversely with the $\Z_2$ fixed plane $\Sigma$.
Namely, we consider the fivebrane wrapped on $\R^4\times C_i$
($i=1,\ldots,n$) where $C_i$ are
two-planes in the fibres of the normal bundle $N$.
Let us choose a rank three orientable subbundle $E$
of the normal bundle $N$
which does not pass through the two-planes $C_1,\ldots,C_n$
except at their origins $p_1,\ldots,p_n$.
The four-cycle we choose is
$\hS=S(E)$, the total space of the two-sphere
bundle in $E$.

Let us measure the flux of $G/2\pi$ through the cycle $S(E)$.
Let $N=L\oplus E$ be the orthogonal decomposition of the normal bundle,
where $L$ is a two-plane bundle over $\Sigma$.
We shall deform the cycle $S(E)$ away from
the origin of the bundle $L$ by using a section $s$
of $L$. It is not in general possible to have a nowhere
vanishing section.
A generic section $s$ has simple zeroes as many as
$\chi(L)$ mod 2, where $\chi(L)=\int_{\Sigma}e(L)$
is the Euler characteristic of
the bundle $L$. (We note that $L$ is orientable since $M^7$ and $E$
are, and we choose one orientation; the other choice
would lead to an opposite answer that is the same mod 2).
Here we have called a zero point $q$ of $s$ a {\it simple zero}
when $s$ induces a diffeomorphism of a neighborhood of $q$ in $\Sigma$
onto a neighborhood of the origin $0$ in the fibre
under a local trivialization of $L$.  
We choose a section $s$ so that
such zero points $q_j$ ($j=1,\ldots,\chi(L)$) are
away from the origin $p_i$
of $C_i$ (intersection points of $C_i$ and $\Sigma$, $i=1,\ldots,n$).
Let us now move the cycle $S(E)$
from the origin of $L$ to the section $s(\Sigma)$ of $L$.
As far as the $S^2$ fibres of $S(E)$
are large enough, the cycle does not pass
through the two-planes $C_i$ nor of course the
$\Z_2$ fixed plane, and hence the flux does not change.
We take a circle in $\Sigma$ encircling each of the zero points $q_j$
of $s$ and intersection points $p_i$ with $C_i$,
and pinch the $S^2$-fibres of the cycle
along the circles as we have done in the case of flat torus.
Then, the cycle splits to several components:
There are $\chi(L)$ four-sphere components surrounding
the $\Z_2$ fixed plane,
$n$ four-sphere components surrounding the $n$ fivebranes,
and one component that surrounds nothing.
As we have seen in the previous subsection, a four-sphere surrounding
the $\Z_2$ fixed plane has flux $1$ mod 2. A sphere surrounding
a fivebrane has of course flux $1$.
Therefore, the total flux through $S(E)$ is $\chi(L)+n$ mod 2,
or equivalently
\beq
2\!\!\int\limits_{S(E)/\Z_2}\!{G\over 2\pi}~=~\chi(L)+n
~~~{\rm mod}~2\,.
\eeq

Now let us evaluate the integral of
$w_4$ over the cycle $S(E)/\Z_2$.
We consider the case where the sphere bundle $S(E)$ over $\Sigma$
is topologically trivial.\footnote{There are only two topological types
of $S^2$ bundles over $\Sigma$ since $\pi_1(SO(3))=\Z_2$.
It is interesting to repeat the computation in the non-trivial one.}
In such a case, $M^7$ is topologically isomorphic to the product
$L\times\R^3$ (we denote the total space of the bundle $L$
again by $L$), and the cycle $S(E)$ and its quotient $S(E)/\Z_2$
corresponds to $\Sigma\times S^2$ in $L\times \R^3$
and $\Sigma\times \RP^2$ in $(L\times \R^3)/\Z_2$
where $\Sigma$ is considered here as the zero section of $L$.
Then, it is easy to see that the tangent bundle of $(M^7-\Sigma)/\Z_2$
restricted to $S(E)/\Z_2$ is isomorphic to the following bundle
over $\Sigma\times \RP^2$:
\beq
\left.T(L\times \R^3-\Sigma\times\{0\})/\Z_2
\right|_{\Sigma\times \RP^2}
~=~~T\Sigma\,\oplus\, (L\otimes m)\,\oplus m^{\oplus 3}\,,
\eeq
where $m$ is the M\"obius line bundle of $\RP^2$, an unorientable
real line bundle whose $w_1(m)$ generates the cohomology ring of
$\RP^2$ over $\Z_2$. Using the well-known properties of the
Stiefel-Whitney class for the sum and tensor product of bundles
\cite{MS},\footnote{The total Stiefel Whiney class
satisfies $w(\xi\oplus\eta)=w(\xi)w(\eta)$
and $w(\xi\otimes\eta)=\prod_{i,j}(1+c_i+d_j)$
if $w(\xi)$ and $w(\eta)$ are expressed formally as
$\prod_i(1+c_i)$ and $\prod_j(1+d_j)$.
See \cite{MS}.}
we see that the fourth Stiefel-Whitney class
of this bundle is given by
$w_4=w_2(L)w_1(m)^2$
which integrates over $\Sigma\times \RP^2$ to $\chi(L)$ mod 2.
Therefore, we have
\beq
\int\limits_{S(E)/\Z_2}w_4~=~\chi(L)~~~{\rm mod}~2.
\eeq
In particular, in the case of flat torus
considered previously,
since the bundle $L$ is trivial
the integral is zero mod 2.

Thus the flux quantiztion condition (\ref{Flux}) is violated
when the number $n$ of fivebranes intersecting with the $\Z_2$
fixed plane is odd. In particular, considering $n=1$ case
we see that a {\it t}-configuration is also inconsistent when the 
$\Z_2$ fixed plane is wrapped on a general Riemann surface $\Sigma$.

This implies that a {\it t}-configuration is
inconsistent even locally.
Namely, the intersection of the fivebrane and the
$\Z_2$ fixed plane is inconsistent by itself
irrespective of what happens elsewhere.
This may sounds not always true if we note that the configuration with
even $n$ satisfies the condition (\ref{Flux}) for the cycle $S(E)/\Z_2$
even in the
case where the fivebranes intersect with $\Sigma$ at distinct points.
However, when $n$ is even, one can avoid a local {\it t}-configuration
by cutting off a disc of each fivebrane around the intersection point
and connecting the boundary cicles
pair-wisely by tubes. These tubes cannot be detected by the cycle
$S(E)/\Z_2$ as far as they are small enough.
When this happens, at each ``intersection point'',
there is a charge flow through the tube, and it is not
a local {\it t}-configuration.
We will see another mechanism of such a
charge flow shortly.
When the number $n$ is odd, pair-off is impossible and a local
{\it t}-configuration is unavoidable.
In any case, all what we have seen implies that a local
{\it t}-configuration is inconsistent.
In order to rigorously show it,
we must consider a cycle that can detect the charge flow along the
$\Z_2$ fixed plane
(such as the flow through thin tubes). Such a cycle must cut through the
$\Z_2$ fixed plane.
Since the $\Z_2$ fixed plane
is a singularity, one cannot consider such a cycle within the low
energy field theory framework. One thing one can do is to
cut off a neighborhood of the $\Z_2$ fixed plane and consider a
space-time with a boundary. In such a case, one needs to know the
generalization of the condition (\ref{Flux}) in the case where
the cycle has a boundary embedded in the boundary of the space-time.
Another thing one may do is to consider the smoothing of the
$\Z_2$ fixed plane, in a way analogous to the smoothing of the
fivebrane as done in \cite{FHMM}.

\subsection*{\sl The $\Z_2$ Fixed Plane 
Screened by a Pair of Fivebranes}

We have seen that it is inconsistent to have the fivebrane intersecting
transversely with the $\Z_2$ fixed plane.
However, the intersection is actually possible
when the $\Z_2$ fixed plane is {\it screened} by a pair of fivebranes,
namely, when there are two fivebranes on top of it.
It can actually arize as a limit of a family of
supersymmetric configurations.
Let us introduce the complex coordinates $z=x^6+ix^{10}$
and $v=x^4+ix^5$.
First, we consider the case where the $\Z_2$ fixed plane is non-compact
in all directions (therefore $z$ is a coordinate of the
complex plane $\C$).
The $\Z_2$ acts as $z\to z$, $v\to -v$ and $x^{7,8,9}\to -x^{7,8,9}$.
We consider a fivebrane wrapped on a $\Z_2$ invariant
holomorphic curve $C$ at $x^{7,8,9}=0$
defined by
\beq
zv^2=\epsilon,
\label{screen}
\eeq
where $\epsilon$ is a parameter.
When $\epsilon$ is non-zero, the curve does not intersect with
the $\Z_2$ fixed plane $v=0$ and the configuration is consistent.
In the limit $\epsilon\to 0$, it degenerates to a curve with
components; one is at $z=0$ and the other is at $v=0$ with
multiplicity two.
The former yields a fivebrane intersecting transversely with the
$\Z_2$ fixed plane and the latter
yields two fivebranes on top of the $\Z_2$ fixed plane.
(Here again, $\epsilon\sim \elel$ is enough
for the configuration to look as described here.)

Thus, a single fivebrane can consistently intersect with
the $\Z_2$ fixed plane which is screened by a pair of
fivebranes.
Doesn't this contradict with what we have seen above when
the $\Z_2$ fixed plane is wrapped on a compact Riemann surface $\Sigma$?
The computation there shows that it is impossible to have a single
fivebrane intersecting with the compact $\Z_2$ fixed plane $\Sigma$,
{\it irrespective of the state of the $\Z_2$ fixed plane}.
In particular, such an intersection is inconsistent
even when a pair of fivebranes are screening
the $\Z_2$ fixed plane $\Sigma$. More generally,
one cannot have the fivebrane
wrapped on odd number of two-planes $C_1,\ldots,C_n$ ($n$ odd)
intersecting transversely with the $\Z_2$ fixed plane $\Sigma$,
no matter whether or not it is screened.

Actually this is not a contradiction.
In fact, when the $\Z_2$ fixed plane is wrapped on a compact surface
$\Sigma$,
it is not possible to find a consistent configuration of the fivebrane
that reduces in some limit to a configuration of odd number
of fivebranes intersecting with the screened $\Sigma$.
Namely we cannot find an analog of (\ref{screen}) when
$x^{6,10}$-directions are comactified.

This can be seen as follows.
We consider the case $M^7=L\times \R^3$
where $L$ is (the total space of)
a holomorphic line bundle over the Riemann surface $\Sigma$.
An analog of (\ref{screen}) would be given by the image $C$ of the
multi-section of the bundle $L$
\beq
\pm\sqrt{\epsilon s(z)}
\label{screen2}
\eeq
where $s$ is a meromorphic section of $L^{\otimes 2}$.
Locally, trivializing $L$ and $L^{\otimes 2}$,
the image $C$ of (\ref{screen2}) can be defined as
$v^2=\epsilon f(z)$ where
$v$ is a complex coordinate of the fibre-direction of $L$
and $f(z)$ is the function associated to the section $s(z)$
with respect to the trivialization.
When the section $s(z)$ has simple poles at $p_1,\ldots,p_n\in\Sigma$,
the $\epsilon\to 0$ limit of the curve $C$ degenerates
and consists of the component $\Sigma$ with multiplicity two
and the components $C_1,\ldots,C_n$ which are fibres
of $L$ at $p_1,\ldots,p_n$.
Namely, in the $\epsilon\to 0$ limit,
the fivebrane wraps twice on the $\Z_2$ fixed plane $\Sigma$
and once on each of $C_1,\ldots,C_n$ which intersect transversely
with the $\Z_2$ fixed plane.
Thus, if $n$ is odd, this appears to contradict with the
flux quantization condition.
However, we must recall the fact that the section $s(z)$ may have
zeroes.
In fact by Riemann-Roch theorem, when $s(z)$ has $n$ poles
it also has $2c_1(L)+n$ zeroes.
Therefore, when $n$ is odd,
there is at least one zero of odd order.
Near such a zero, say simple zero, the section
$s(z)$ or its associated function behaves as $f(z)\sim z$,
and the equation defining the curve $C$ looks like
\beq
v^2\sim \epsilon z.
\eeq
Thus, before the $\epsilon\to 0$ limit,
 we see that the curve $C$ intersects transversely with
the $\Z_2$ fixed plane, which is a {\it t}-configuration.
When the order of zero is higher but odd,
we can similarly see that odd number of sheets of the fivebrane
intersect with
the $\Z_2$ fixed plane at a single point.
In any case, if $n$ is odd,
the configuration was inconsistent from the
start (i.e. before the $\epsilon\to 0$ limit).
Thus, the apparent contradiction is resolved.

If $n$ is even in the above discussion,
it is possible to find a section $s$ of $L^{\otimes 2}$ where
all the zero points are double zeroes.
Then, the configuration is consistent and the $\epsilon\to 0$
limit leads to a configuration
of $n$ fivebranes intersecting with the $\Z_2$ fixed plane
screened by a pair of fivebranes.
In the particular case where $L$ is the holomorphic cotangent bundle of
$\Sigma$ (this is the case where the total space
$L$ admits a Ricci-flat K\"ahler
metric, e.g. the Taub-NUT metric),
this can be done without breaking supersymmetry.
This is because the number of independent meromorphic
sections with simple poles at given points $p_i$
is $2c_1(L)+n+(1-g)=3(g-1)+n$ where $g$ is the genus of $\Sigma$
while we only need to tune $(2c_1(L)+n)/2=2(g-1)+n/2$ parameters
corresponding to the position of half of the zero points.
In the case of $g=0$ and $n=2$,
there is no such section and therefore
we cannot obtain the configuration in this way.
However, there is no obstruction to having such a configuration.
Indeed we will need such a configuration in a later discussion,
and the fact that it cannot be deformed to a smooth configuration as
(\ref{screen2}) turns out to be important.

{\bf To summarize},
{\it {\it t}-configuration is inconsistent.
$n$ fivebranes intersecting with the $\Z_2$ fixed plane
at the same point is consistent if and only if $n$ is even.
A single fivebrane (and any number of them) can intersect with
an infinite $\Z_2$-fixed plane which is screened by
a pair of fivebranes on top of it.
When the $\Z_2$ fixed plane is compactified in the directions
transverse to the fivebrane,
odd number of fivebranes can never intersect with
it even if it is screened by the fivebranes,
but there is no obstruction for even number of fivebranes to
intersect with it if it is screened.}

\section{Compactification}

By compactifying several directions,
we can derive from what we have learned
in the previous section
some consistency conditions for
various intersection
of branes and orientifold plane in string theory.
As we will see,
some of them are related to known or conjectured
consistency condition.

\subsection*{\sl Intersection of NS5-brane and Orientifold Plane}

We consider \MT on the
space-time $\R^5\times \R^5/\Z_2\times S^1$.
As in the previous section, we coordinatize the space-time
by $x^0,\ldots,x^{10}$ where $\Z_2$ acts as the sign flip of
$x^{4,5,7,8,9}$. $x^{10}$ is now a periodic coordinate of
period $2\pi$ which we regard as the eleventh direction.
In the small radius limit of $S^1$, we can consider it as
Type IIA orientifold on $\R^5/\Z_2$ where there is an
orientifold four-plane (O4-plane) at $x^{4,5,7,8,9}=0$.
Since the $\Z_2$ fixed plane $\R^5\times S^1$ has fivebrane charge
$-1$, this O4-plane has D4-brane charge $-1$ and therefore
can be identified with the O4-plane of $SO$-type.

%The consistency conditions for the transverse intersection
%of fivebrane and $\Z_2$ fixed plane in \MT translate
%to similar conditions for the intersection of NS5-brane and
%O4-plane with a possible
%modification of the interpretation.

Suppose we have a single NS5-brane at $x^{6,7,8,9}=0$
transversely intersecting with the $SO$-type O4-plane.
This is realized as \MT on
$\R^5\times \R^5/\Z_2\times S^1$
with a single fivebrane at $x^{6,7,8,9}=0$ and at
some point in the $S^1$ direction.
The fivebrane intersects transversely
with the $\Z_2$ fixed point set,
and therefore the configuration is inconsistent.
Similarly, if there are several NS5-branes intersecting with the
O4-plane at the same point, it is consistent if and only if the
number of NS5-branes is even.

In the previous section, we have shown that
a single fivebrane can transversely intersect
with the $\Z_2$ fixed plane provided there are two fivebranes on top
of the $\Z_2$ fixed plane.
This does not mean that an NS5-brane can intersect
with the $SO$-type O4-plane with two D4-branes on top of it. 
\footnote{As another example showing that the condition in \MT
does not necessarily have a direct translation in string theory,  
we point out that, even though
a single \MT fivebrane cannot be on top of
$\R^5/\Z_2$ fixed point, a single D4-brane
can be on top of O4-plane of $SO$-type.
In section 5 and 6,
we will give the \MT realization of the latter configuration.}
Rather, the correct interpretation of this is the following.
Note first that the NS5-brane divides the O4 plane to two parts ---
the left part $x^6<0$ and the right part $x^6>0$.
Then, the O4-plane is of $Sp$-type
on one part while it is $SO$-type O4-plane with
a pair of D4-branes on the other part.
This can be seen as follows.
Let us introduce the complex coordinates $t=\e^{-(x^6+ix^{10})}$
of the cylinder in the $x^{6,10}$ directions and $v=x^4+ix^5$
of the $x^{4,5}$-plane. The configuration can be described by
the equation $v^2(t-1)=0$. This can be deformed as
\beq
v^2(t-1)+m^2=0,
\label{defo}
\eeq
(or as $v^2(t-1)=m^2 t$ which is related to the above
by $t\to t^{-1}$)
without having a {\it t}-configuration.
In the left infinity $t\to\infty$, the fivebrane wraps twice
on the cylinder at $v=0$ whereas it splits to
$v=m$ and $-m$ on the right infinity $t\to 0$.
Since $m$ can be varied, we interpret the right part as
the $SO$-type O4-plane with a pair of D4-branes.
Since the fivebrane in the left part is fixed at $v=0$
and adds the D4-brane charge $+2$ to the charge $-1$ from the 
$\Z_2$ fixed plane, yielding the total charge $+1$,
it is most natural to interpret the left part
as O4-plane of $Sp$-type.

We can get rid of the pair of D4-branes on the right part
by sending $m\to\infty$,
and we obatin a configuration where an NS5-brane separates
the orientifold-plane into $Sp$-type O4-plane and $SO$-type
O4-plane.
This is described in \MT by the fivebrane wrapping the curve $v^2t=1$
which can be obtained from (\ref{defo}) by taking
the $m\to\infty$ limit with a suitable translation in the
$x^6$-direction.
We will study more about the intersection of NS5-brane and O4-plane
in section 7.

Compactification and T-duality
on some of the $x^{0,1,2,3}$ directions
yields a configuration where the NS5-brane divides the
orientifold $p$-plane ($p\leq 4$) into two parts
--- $Sp$-type O$p$-plane on one part and $SO$-type
O$p$-plane on the other.

\subsection*{\sl Fivebrane in Type I String Theory}

We next consider \MT on $\R^5\times T^5/\Z_2\times S^1$.
Regarding the last $S^1$ as the eleventh direction which
we take to be small,
we consider this as Type IIA orientifold on $T^5/\Z_2$
which is T-dual to Type I string theory on $\tilde T^5$.
We coordinatize the space by $x^{0,1,\ldots,10}$
where $x^{4,5,7,8,9}$ and $x^{10}$ are the coordinates
of period $2\pi$ parametrizing $T^5$ and the eleventh $S^1$
respectively.
Suppose we have a longitudinal fivebrane at $x^{3,6,7,8,9}=0$
which spans the $x^{0,1,2,4,5,10}$ directions.
Since it intersects with a $\Z_2$ fixed plane,
this is a {\it t}-configuration.
This fivebrane is a D4-brane in Type IIA orientifold
which is T-dual to a single D5-brane in Type I side.
Therefore, we conclude that a single D5-brane in Type I string theory
is inconsistent.
It is known that open strings ending on
D5-branes in Type I string theory
must carry $Sp(1)$ or $Sp(N)$ Chan-Paton factor
and therefore two D5-branes move together as a unit
(where the number is counted in the double cover of
Type IIB orientifold)
\cite{GP}.\footnote{One of the argument showing this as mentioned in
\cite{GP} is the Dirac quantization condition for D5 and D1 branes.
This is closely related to, or may
be mapped to, the consideration in section
2.}
This looks consistent with the condition we obtained.

Actually, there is a subtlety in the above argument.
There are 16 pairs of \MT fivebranes which are at points of $T^5$
and are parallel to the 32 $\Z_2$ fixed planes \cite{DM,orbW}.
Recall again that, if a $\Z_2$ fixed plane is with two
parallel fivebranes on top of it, a fivebrane can
transversely intersect with it. Since the fivebrane in the above
situation passes through the four $\Z_2$ fixed planes
at $x^{7,8,9}=0$,
the configuration is consistent if each of these four fixed planes
is screened by a pair of fivebranes on top of it.
However, in order to obtain ten dimensional
Type I string theory with $SO(32)$
unbroken gauge symmetry by decompactifying $\tilde T^5$,
all of the 16 pairs must be at one
fixed plane. But this is impossible without having a
{\it t}-configuration. This is a more precise way to relate
our condition to that of \cite{GP}.

In section 2, we found that it is consistent to have
two fivebranes intersecting with the
$\Z_2$ fixed plane at the same point,
even if the two are not necessarily parallel.
It may appear that we can conclude from this that
two non-parallel fivebranes is possible in
Type I string
theory, which contradicts the condition of \cite{GP}.
The solution of this puzzle is that a {\it t}-configuration cannot
be avoided if there are only two
fivebranes, as can be seen as follows.
Suppose one fivebrane is at $x^{3,6,7,8,9}=0$ (spanning
$x^{0,1,2,4,5,10}$) and the other is at
$x^{3,4,5,6,7}=0$ spanning ($x^{0,1,2,8,9,10}$).
They do avoid a {\it t}-configuration at $x^{4,5,7,8,9}=0$. But
the first fivebrane passes through the $\Z_2$ fixed plane at
$(x^4,x^5)=(0,\pi),(\pi,0),(\pi,\pi)$ by itself (and similarly
for the second fivebrane), and they are {\it t}-configurations.
In order to avoid {\it t}-configuration at all the
32 $\Z_2$ fixed plane, we must have a fivebrane
spanning $x^{0,1,2,4,5,10}$ at each  of
the four fixed points in the
$x^{8,9}$-directions and other four spanning $x^{0,1,2,8,9,10}$
at the four fixed points in the $x^{4,5}$ directions.
After T-duality, we have a block of
four D5-branes on top of each other
with another block of four D5-branes,
which is of course consistent.

\subsection*{\sl Other Intersection of D-branes and
Orientifold Plane}

Starting
from a {\it t}-configuration with the fivebrane wrapped on the
eleventh direction,
we can derive similar consistency conditions
by compactifying
various other directions of $\R^5\times \R^5/\Z_2$.
Namely, in Type II string theory,
it is impossible to have
a single D$p$-brane and an $SO$-type O$p^{\prime}$-plane
intersecting transversely in a four-dimensional factor
of the ten-dimensional space-time.\footnote{
The case of a D4-brane intersecting with $SO$-type O6-plane
was conjectured to be impossible in \cite{EGKT,Caba}.}
In section 7, we will study the intersection of D6-brane and
various type of O4-planes in detail and determine what kind of
intersection is consistent.

When the $\Z_2$ fixed plane is screened by a pair of fivebranes,
the fivebrane can consistently intersect with it.
If we compactify a direction parallel to both
the fivebrane and the $\Z_2$ fixed plane on a circle
and consider it as
the eleventh direction, the fivebrane is of course
identified as a D4-brane.
The screened $\Z_2$ fixed plane appears to
be interpreted as an $Sp$-type O4-plane as in the case of NS5-brane
intersecting with the O4-plane.
However, unlike in that case, there is a supersymmetry preserving
deformation of the configuration (turning on $\epsilon$
in (\ref{screen})) that has apparently no counterpart in $Sp$-type
O4-plane. Thus, we simply interpret the configuration as
the $\epsilon\to 0$ limit of D4-brane wrapped on the curve
(\ref{screen}) in the presence of the $SO$-type O4-plane.

\section{Brane Configuration for $N=2$ SQCD with Symplectic
Gauge Groups}

In \cite{DSB}, in addition to the hypothesis that a
{\it t}-configuration is not supersymmetric (which
we have proved in section 2), a generalized hypothesis
in the presence of D6-branes was proposed.
This was to obtain correctly the dimension
of the Higgs branch of $Sp(N_c)$ SQCD in four dimensions by
the corresponding brane configuration.
Although it works for the purpose of Higgs branch counting,
the configurations contain {\it t}-configurations
 as we will see, and are inconsistent by themselves.
In this section, we modify the brane configuration and show that
the Higgs branch counting works
perfectly without any assumption
provided we use the consistency condition which we derived
in section 2 from the flux quantization condition (\ref{Flux}).

We first describe the eleven-dimensional space-time.
It is the quotient of $\R^7$ times a Taub-NUT space
by a certain $\Z_2$ action.
Let us choose the time and space
coordinates $x^0,x^1,\ldots,x^{10}$ so that $x^{4,5,6,10}$
parametrize the Taub-NUT space where $x^{10}$ is the periodic
coordinate of the eleventh direction.
The Taub-NUT geometry in \MT corresponds to D6-branes
in Type IIA string theory which can be used to realize quark
multiplets.
It can be considered as the circle bundle over $\R^3$
(circle and $\R^3$ parametrized by $x^{10}$ and
$\vec{x}=x^{4,5,6}$ respectively)
where the size of the circle shrinks at
the location $\vec{x}=\vec{x}_i$ of the D6-branes ($i=1,\ldots, 2N_f$).
We only consider the essential case where all the quarks are massless.
Then, $\vec{x}_i$ are aligned as $x^6_i<x^6_{i+1}$,
$x^4_i=x^5_i=0$.
We choose a complex structure such that $v\propto x^4+ix^5$ is a
holomorphic coordinate. As other coordinates, we can take $y$ and $x$
which are related to each other by
\beq
xy=v^{2N_f}.
\eeq
These are related to the real coordinates by
$y=\e^{-(x^6/R+ix^{10})}f(\vec{x})$ and
$x=\e^{x^6/R+ix^{10}}g(\vec{x})$
where $f$ and $g$ are certain functions that vanish at
the position $\vec{x}=\vec{x}_i$
of the D6-branes (see \cite{NOYYTN}).
The description in terms of $y,x,v$ breaks down near the
D6-branes. As long as the sixbranes are separated in the $x^6$
direction, $x^6_i\ne x^6_j$,
 the space is smooth and is described by introducing
one coordinate system $(y_i,x_i)$ in a neighborhood
of each of them. 
These coordinate systems are related to each other 
by $(y_{i-1},x_{i-1})=(y_i^2x_i,y_i^{-1})$ and are related to
$(y,x,v)$ as
$y=y_i^ix_i^{i-1}$,
$x=y_i^{2N_f-i} x_i^{2N_f+1-i}$,
and $v=y_ix_i$.
There are
$2N_f-1$ $\CP^1$ cycles $C_i$
defined by $y_i=x_{i+1}=0$ for $i=1,\ldots,2N_f-1$.
The cycle $C_i$ can also be defined as the fibres over the straight
segment stretched between the two D6-branes
at $\vec{x}_i$ and $\vec{x}_{i+1}$
when the space is considered as the circle bundle over
the $\vec{x}$-space.

The $\Z_2$ action is given by the sign flip of $x^{4,5,7,8,9}$.
The action on the complex coordinates
of the Taub-NUT space is $(x,y,v)\to (x,y,-v)$, and  it follows
from this that the action on the local coordinates $(y_i,x_i)$
is
\beq
y_i\to (-1)^{i-1}y_i,~~~x_i\to (-1)^i x_i.
\label{Z2i}
\eeq
Since $\Z_2$ also acts on the coordinates $x^7,x^8,x^9$ as
$x^{7,8,9}\to-x^{7,8,9}$ at the same time,
the $\Z_2$ fixed points must be at $x^{7,8,9}=0$.
From (\ref{Z2i}), we see
that the $\CP^1$ cycles $C_{2i}$ at $x^{7,8,9}=0$
are point-wisely $\Z_2$ invariant
but the $\Z_2$ acts on the other cycles
$C_{2i-1}$ as $\pi$-rotation.
The semi-infinite cycles $\{x_1=0\}$ ($y$-axis)
and $\{y_{2N_f}=0\}$ ($x$-axis) at $x^{7,8,9}=0$
are also point-wisely $\Z_2$ fixed.
\footnote{There is one more ingredient which is
important for the low energy physics in the Coulomb branch
though it is not the focus of the present paper.
It is the holomorphic two-form $\Omega=\dd v\dd y/y=-\dd v\dd x/x$
of the Taub NUT space. Since
$\Omega=\dd y_i\dd x_i$ in the $i$-th patch, it is nowhere vanishing
and flips sign under the $\Z_2$ action. 
A BPS state is associated with a minimal-area membrane ending on
the fivebrane whose orientation flips under the $\Z_2$ action
(because of the parity transformation law $C\to -C$ of the three-form
potential), and its mass is given by the absolute value of
the integral of $\Omega$ on it (which is well-defined
since $\Omega$ also flips sign).
Likewise the set of cycles to determine the prepotential
are chosen from anti-invariant cycles.}

The fivebrane wraps on the curve given by
\beq
x+y={v^2\over
\Lambda^{2N_c+2-N_f}}\prod_{a=1}^{N_c}(v^2-\phi_a^2)\,,
\label{Spcurve}
\eeq
where $\pm \phi_a$ are the eigenvalues of the adjoint
chiral superfield and $\Lambda$ is the dynamical scale of the theory.
This form is determined by the asymptotic behaviour at $v\to\infty$,
the $\Z_2$ invariance, and the requirement that the curve does not
intersect transversely with the semi-infinite $\Z_2$ fixed planes
--- $y$ and $x$-axis.
Let us first look at a generic point of the Coulomb branch where
all $\phi_a$ are not zero. As in \cite{W2,HOV,HOO},
we see that the fivebrane wraps the $\CP^1$ cycles
$C_1,C_2,\ldots,C_{2N_f-2},C_{2N_f-1}$ with multiplicity
$1,2,\ldots,2,1$. Also, an infinite component intersects
$C_2$ at one point and $C_{2N_f-2}$ at one point.
Only from the requirement of the $\Z_2$ invariance of the configuration,
there is nothing to prevent the two components wrapped on
$C_2,\ldots,C_{2N_f-2}$, to be separated in the $x^{7,8,9}$-directions,
which would correspond a Higgs branch
of the worldvolume theory.
But this does not agree with the
field theory knowledge; In $N=2$ $Sp(N_c)$ SQCD, no Higgs branch
emanates from a generic point of the Coulomb branch.
Even if we require that the componets wrapping
$C_2$ and $C_{2N_f-2}$ to be at $x^{7,8,9}=0$ (to avoid
the transverse intersection of the infinite component with the bare
$\Z_2$ fixed plane), there are still many compoents that can move.

In order to reconcile the discrepancy,
in \cite{DSB}, the following hypothesis was proposed:
The configuration is supersymmetric
only if the intersection points of the infinite component
with the $\CP^1$ cycles $C_2$ and $C_{2N_f-2}$
are connected by a series of $\CP^1$ components.
%Indeed, under this constraint
%there is no $\Z_2$ invariant way to move the components wrapped on
%$C_i$ away from $x^{7,8,9}=0$. 
It was shown in \cite{DSB} that this rule works in
all the cases as far as the Higgs branch counting is concerned.

However, the configuration considered above contains
a {\it t}-configuration and is actually inconsistent by itself.
Let us look at the point $y_1=x_1=0$, the position of the first
D6-brane. The fivebrane wraps once on the cycle $C_1$ described by
$y_1=0$ which
transversely intersects the $y$-axis given by $x_1=0$.
Since $\Z_2$ acts on the coordinates $(y_1,x_1,x^{7,8,9})$
as $(y_1,-x_1,-x^{7,8,9})$, $y$-axis is a
$\Z_2$-fixed plane and therefore this is a {\it t}-configuration.
This kind of local {\it t}-configuration appears for arbitrary
values of $\phi_a$.

The problem appears to be solved if we remove the component wrapping
$C_1$ and $C_{2N_f-1}$. Indeed, if we do this, the {\it t}-configuration
is removed. However, there are odd number of fivebranes intersecting
with the compact $\Z_2$ fixed plane $C_2$ (or $C_{2N_f-2}$) --- the
single infinite component and the two components wrapped on
$C_3$ (or on $C_{2N_f-3}$). This violates the consistency condition
derived in section 2; the flux quantization condition requires that
the number of fivebranes intersecting with such a $\Z_2$ fixed plane
must be even, even if the fixed plane is screened by a pair of
fivebranes. This new problem can be solved by removing
a single component from each of the $\CP^1$ cycles $C_3$ and
$C_{2N_f-3}$. Then, the same kind of problem
arizes at the compact $\Z_2$ fixed
planes $C_4$ and $C_{2N_f-4}$. Continuing this process we finaly
obtain a consistent configuration which is described as follows:

In addition to the infinite component,
the fivebrane
wraps on the $\CP^1$ cycles $C_1,\ldots,
C_{2N_f-1}$  with multiplicity
$0$ for $C_1,C_{2N_f-1}$, $1$ for other
$C_{2i-1}$, $2$ for $C_{2i}$
\begin{figure}[htb]
\begin{center}
\epsfxsize=4.5in\leavevmode\epsfbox{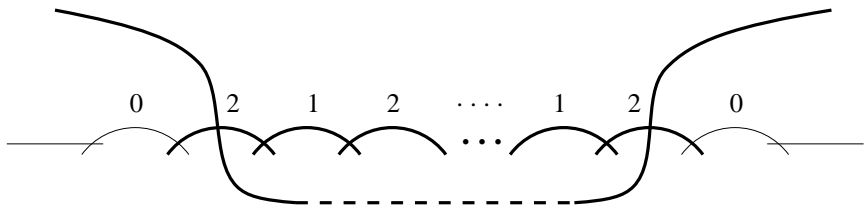}
\end{center}
\caption{The Corrected Curve (a Generic Point of the Coulomb Branch)} 
\label{Coul}
\end{figure}
(see Figure \ref{Coul}).
When all the $\CP^1$ components are at $x^{7,8,9}=0$,
this is a consistent configuration: the total number of fivebrane
intersecting with the $\Z_2$ fixed cycle $C_{2i}$ is even,
which is required from the flux quantization condition.
Also, the number of components wrapping each of
these fixed cycles $C_{2i}$ is even, and another consistency
requirement is satisfied.
Furthermore, in order to avoid the {\it t}-configuration,
the two components wrapping each of the cycles $C_{2i}$ cannot be
separated in the $x^{7,8,9}$-direction. This corresponds to the absence
of Higgs branch at a generic point of the Coulomb branch,
which agrees with the field theory knowledge.

Thus,
we propose a modification of the brane configuration so that
any problem does not arize:
{\it We remove one component
of the fivebrane wrapping each of the cycles $C_{2i-1}$
from what one would get by the equation (\ref{Spcurve}).}
This is possible 
for an arbitrary values of $\phi_a$
since the fivebrane (\ref{Spcurve}) always wraps
on these cycles at least once.

Let us consider the case where $r$ of $\phi_a$'s vanish ($r$
sufficiently small). The equation
(\ref{Spcurve}) would imply that there are
$\CP^1$ components wrapped on
$C_1,\ldots,C_{2N_f-1}$ with multiplicity
$1,2,3,\ldots,2r+1,2r+2,\ldots,2r+2,2r+1,\ldots,3,2,1$.
But this is modified as follows: the number of wrapped fivebrane
is $2[i/2]$ for $C_i$ and $C_{2N_f-i}$ with $i<2r+2$;
$2r+1$ for $C_{2i+1}$ with $r<i<N_f-r-1$;
$2r+2$ for $C_{2i}$ with $r<i<N_f-r$.
The total number of fivebrane
intersecting with each of the $\Z_2$ fixed cycles $C_{2i}$
is even (For $i\leq r$ or $i\geq N_f-r$ this is trivial as the
multiplicities at $C_{2i-1}$ and $C_{2i+1}$ are both even.
For $r+1<i<N_f-r-1$. they are both odd and the sum is again even.
For $C_{2r+2}$ or $C_{2N_f-2r-2}$, the sum of mulitiplicities
of the nighboring cycles is odd, but the infinite component
intersects with it at one point.).
In addition, the multiplicity at these $\Z_2$ fixed cycles $C_{2i}$
is even. Therefore the configuration
is consistent.
Also, in order to avoid a {\it t}-configuration,
two of the $\CP^1$ components wrapping $C_{2i}$ with $r<i<N_f-r$
must be at $x^{7,8,9}=0$, and there is no other constraint.
It is easy to see that the number of $\Z_2$ invariant
deformations of
$\CP^1$ components subject to this constraint
agrees with the dimension $2rN_f-(2r^2+r)$
of the $r$-th Higgs branch.
It is also easy to see that the subtlety associated with
the $r=[N_f/2]$ Higgs branch \cite{APSh,DSB}
is correctly captured by applying the consistency
condition to the modified curve.

Starting with this $N=2$ configuration, we can obtain the
configuration for $N=1$ SQCD which will not be described in this paper.
Here we only mention that one can also obatin results
consistent with field theory from the modified configuration.
See \cite{H}.

\section{Brane Configuration for $N=2$ SQCD with Orthogonal
Gauge Groups}

In this section,
we construct brane configuration corresponding to $N=2$ theories
with orthogonal gauge groups.
No satisfactory description of these theories
including the Higgs branch has been known in the past.
We first show that the na\"\i ve candidate of the
configuration for
$SO(N_c)$ SQCD satisfies the consistency requirement for
even $N_c$ but not for odd $N_c$.
In the process of realizing $SO({\rm odd})$ gauge theories,
we find a new \MT realization of orientifold four-planes.
We also encounter a phenomenon where fivebranes are created when two
coincident sixbranes are separated while intersecting with the
orientifold four-plane.

\subsection{$SO({\rm even})$ Gauge Groups}

The configuration corresponding to
$N=2$ $SO(N_c)$ SQCD, even $N_c$,
with $N_f$ flavors ($N_f$ hypermultiplets in the vector representation)
can be constructed in the eleven-dimensional space-time
which is the same as in the case of $Sp(N_c)$ SQCD with $N_f$ flavors.
We thus follow the notation used in section 4.

At a point on the Coulomb branch, the fivebrane wraps on the curve
described by \cite{LLL}
\beq
v^2(x+y)={1\over \Lambda^{N_c-2-N_f}}
\prod_{a=1}^{N_c/2}(v^2-\phi_a^2)
\label{SWeven}
\eeq
where $\pm \phi_a$ are the eigenvalues of the adjoint chiral
superfield and $\Lambda$ is the dynamical scale
of the theory.
The $v^2$ factor of the left hand side shows that,
in the asymptotic region $x^6\to\pm\infty$,
the fivebrane wraps twice on the
$\Z_2$-fixed cylinder $x^{4,5,7,8,9}=0$, $0\leq x^{10}<2\pi$.
In the Type IIA limit, the $\Z_2$ fixed plane in this region
has D4-brane charge $-1+2=+1$
and can be interpreted as the orientifold four-plane of $Sp$-type.

Higgs branches emanate from the submanifolds of the Coulomb branch
where some of $\phi_a$'s vanish.\footnote{
Unlike in the $Sp$ case (see \cite{APSh,DSB}),
the theory at the root is always
infra-red  free and therefore the semi-classical intuition
about the location of the root is correct.}
When $r$ of them vanish ($r\leq {N_c\over 2}, {N_f\over 2}$),
the theory at the root is $SO(2r)$
with $N_f$ flavors, and we expect a Higgs branch of
quaternionic dimension $2rN_f-2r(2r-1)/2$.
The curve (\ref{SWeven}) at such a point is singular and contains
three infinite components --- the left, middle and right components:
The left component wraps twice on the $y$-axis ($x^{4,5,7,8,9}=0$,
$x^6<$ any $x^6_i$, $0\leq x^{10}<2\pi$),
the right wraps
twice on the $x$-axis ($x^{4,5,7,8,9}=0$,
$x^6>$ any $x^6_i$, $0\leq x^{10}<2\pi$),
and the middle wraps on a curve $C$ extending to $v\to\infty$.
The fivebrane also wraps on the $\CP^1$ cycles
$C_1,\ldots,C_{2N_f-1}$ with multiplicities
$3,4,\ldots,2r-1, 2r,\ldots,2r,2r-1,\ldots,4,3$.
The component $C$ intersects
transversely with $C_{2r-2}$
and $C_{2N_f-2r+2}$ at one point each (if $r>1$).
The total number of fivebranes intersecting with the $\Z_2$ fixed
plane $C_{2i}$ is even
(The number of fivebranes wrapping the neighboring cycles
$C_{2i-1}$ and $C_{2i+1}$ are both odd for $i<r-1$ or $i>N_f-r+1$;
both even for $r\geq i\leq N_f-r$, and therefore the sum
is even for these cases.
For $C_{2r-2}$ or $C_{2N_f-2r+2}$, the sum of mulitiplicities
of the nighboring cycles is odd, but the infinite component $C$
intersects with it at one point.).
In addition, the fivebrane wraps on these $\Z_2$ fixed cycles
$C_{2i}$ with even number of times.
Therefore the configuration is consistent.
Since the number of fivebranes wrapping 
$C_{2i+1}$ is odd for $i\leq r-1$ or $i\geq N_f-r+1$,
at least one component wrapping these cycles must be at
$x^{7,8,9}=0$.
Therefore, in order to avoid {\it t}-configuration,
at least one pair of components wrapping each of $C_i$ and $C_{2N_f-i}$
with $i=1,\ldots,2r-2$ must be at $x^{7,8,9}=0$.
The number of $\Z_2$ invariant deformations of the location of
$\CP^1$ components subject to this constraint agrees with the expected
dimension of the Higgs branch.

\subsection{$SO({\rm odd})$ Gauge Groups}

\subsection*{\sl A Problem}

It appears that the curve for $SO(N_c)$ gauge theory
with $N_c$ odd can be obtained by replacing the right hand side
of (\ref{SWeven}) by $v\prod_{a=1}^{[N_c/2]}(v^2-\phi_a^2)$
with a modification of the $\Z_2$ action.
The curve thus obtained is indeed the same as the Seiberg-Witten curve
of the theory.
However,
the dimension of the Higgs branch does not agree with the number of
motion of the $\CP^1$ components.
This can be easily seen by noting that
an $N_f$ dimensional Higgs branch emanates at a generic point of the
Coulomb branch \cite{APSh} in $SO({\rm odd})$ theories, while
for generic values of $\phi_a$ the fivebrane thus obtained wraps only
once on each of the $\CP^1$ cycles and hence they are fixed at
$x^{7,8,9}=0$.

There is actually a more fundamental reason why this cannot be
the correct fivebrane configuration.
For an arbitrary value of $\phi_a$, there is a $\Z_2$ fixed
cycle $C_{2i}$ on which the fivebrane wraps odd number of times.
The fivebrane wrapping odd number of times on a
$\CP^1$ cycle which is point-wisely $\Z_2$ fixed is
locally the same as odd number of fivebranes stuck at the orbifold
point of $\R^5/\Z_2$. This configuration is forbidden since it
does not satisfy the flux quantization
condition \cite{Wflux} (see section 2).
Therefore, the configuration given above is inconsistent.

\subsection*{\sl $SO({\rm odd})$ from $SO({\rm even})$ by Higgsing}

In field theory, one can reduce the number of colors
(and flavors) by Higgsing the quarks.
As suggetsed in \cite{GK},
one can use this fact to obtain the right configuration
for $SO({\rm odd})$ theories from the configuration for
$SO({\rm even})$ theories.
%In order to see what is the precise analog of the required Higgsing
%in the brane picture, we start with small values of $N_f$.

We first consider obtaining $SO(N_c)$ pure $N=2$
Yang-Mills theory ($N_c$ odd, $N_f=0$)
from $SO(N_c+1)$ theory with a single flavor.
This can be done by setting $\phi_{{N_c+1\over 2}}=0$,
turning the quark vev $Q^{i=1,2}_{a=N_c+1}$
and sending them to infinity.
The corresponding procedure in the brane picture is obvious.
At $\phi_{{N_c+1\over 2}}=0$, the curve of $SO(N_c+1)$
theory is given by
$v^2(x+y)=v^2\prod_{a=1}^{[N_c/2]}(v^2-\phi_a^2)/\Lambda^{(N_c+1)-2-1}$
and wraps twice on the unique $\CP^1$ cycle of the $A_1$ type
Taub-NUT space $yx=v^2$.
The quark vevs correspond to the position of
the pair of $\CP^1$ components in the $x^{7,8,9}$-directions
and hence the configuration for $SO(N_c)$ Yang-Mills theory
can be obtained by sending them to $x^{7,8,9}\to\infty$.
At the same time, we can send the two sixbranes to infinity
$x^6\to\pm\infty$; one to the left $x^6_1\to -\infty$
and one to the right $x^6_2\to +\infty$.
Then, the $\CP^1$ cycle becomes infinitely elongated
and it is appropriate to use the coordinates
$\tilde{x}=v^{-1}x$ and $\tilde{y}=v^{-1}y$ related by
$\tilde{y}\tilde{x}=1$.
The expression of the curve is then
\beq
v(\tilde{x}+\tilde{y})={1\over\Lambda^{N_c-2-N_f}}
\prod_{a=1}^{[N_c/2]}(v^2-\phi_a^2),
\label{oddYM}
\eeq
where $N_f=0$ in the present case.
The overall factor $v^2$ has dropped off because the two $\CP^1$
components has been sent to infinity.

By definition, the $\Z_2$ acts on the new coordinates as
$\tilde{y}\to-\tilde{y}$, $\tilde{x}\to -\tilde{x}$.
In other words, the space-time obtained by sending the
sixbranes to $x^6=\pm\infty$ is $\R^5\times(\R^5\times S^1)/\Z_2$
where the $\Z_2$ now acts on $\R^5\times S^1$
with coordinates $x^{4,5,7,8,9}$ and $x^{10}$ as
\beq
\begin{array}{l}
x^{4,5,7,8,9}\to-x^{4,5,7,8,9}\\[0.1cm]
x^{10}\to x^{10}+\pi.
\end{array}
\label{newZ2}
\eeq
Note that there is no fixed point of this $\Z_2$ action.
In particular, there is no extra fivebrane charge that would be
associated with the $\R^5/\Z_2$ fixed point \cite{DM,orbW}.
In the weakly coupled Type IIA limit where
the size of $S^1$ shrinks to zero, the $\Z_2$-invariant
cylinder $x^{4,5,7,8,9}=0$ looks like a $\Z_2$
fixed plane in ten-dimensional space-time but it carries no D4-brane
charge by itself.

In the asymptotic region $|x^6|\to\infty$,
the fivebrane wraps once
on this $\Z_2$-invaraint cylinder
because of the factor of $v$ in the left hand side
of (\ref{oddYM}).
Therefore the $\Z_2$ ``fixed plane''
$x^{4,5,7,8,9}=0$ carries D4-brane charge $+1$
in the large $x^6$ region.
This can be interpreted as the O4-plane of $Sp$-type.
In the middle region $\tilde{x}\sim\tilde{y}$,
there are ${N_c-1\over 2}$ pairs of D4-branes parallel to
this $\Z_2$ ``fixed'' plane of chrage zero.
We interpret this $\Z_2$ ``fixed'' plane in this region as
O4-plane of $SO$-type with a single D4-brane stuck on it.

\subsection*{\sl A Transition with Fivebrane Creation}

Let us next consider obtaining $N=2$ $SO(N_c)$ SQCD ($N_c$ odd)
with $N_f=1$ from $SO(N_c+1)$ theory with two flavors.
In field theory, this can be done by setting $\phi_{N_c+1\over 2}=0$
and sending some of the quark vevs
%$Q^{i=1,2,3,4}_{a=N_c+1}$
to infinity, throwing away two quaternionic degrees of freedom.
In the brane side, the fivebrane for $SO(N_c+1)$ theory
with $\phi_{N_c+1\over 2}=0$
wraps twice on each of the three $\CP^1$ cycles $C_1,C_2,C_3$.
The decoupling is done by sending two pairs of $\CP^1$
components to infinity.
Here we consider sending the pairs wrapping $C_1$ and $C_3$
to infinity.
(We do not consider here the other two options which might lead to
other realization of $SO({\rm odd})$ theories.)
If we send also the leftmost and rightmost sixbranes
to $x^6\to-\infty$ and $+\infty$ at the same time,
the cycles $C_1$ and $C_3$ becomes infinitely elongated
and it is appropriate to use the coordinates
$\tilde{y}=v^{-1}y$ and $\tilde{x}=v^{-1}x$. These are related
to each other by $\tilde{y}\tilde{x}=v^2$
the $\Z_2$ acts on them as $\tilde{y}\to -\tilde{y}$
and $\tilde{x}\to -\tilde{x}$.
The expression of the curve is then given by
(\ref{oddYM}) in which $N_f=1$ in this case.
Here we should keep in mind that
we are retaining the pair of $\CP^1$ components wrapping $C_2$,
although we have dropped the $v^2$ factor from the original
expression for the $SO(N_c+1)$ theory.

\begin{figure}[htb]
\begin{center}
\epsfxsize=5.2in\leavevmode\epsfbox{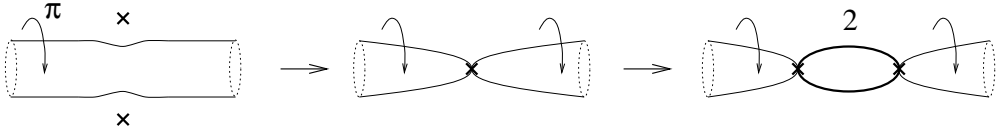}
\end{center}
\caption{Creation of Fivebrane}
\label{creation}
{\small  A pair of sixbranes approach the cylinder on which
$\Z_2$ acts as $\pi$-rotation (left). When they coincide, the
cylinder is pinched (middle). When they are separated,
a $\CP^1$ cycle which is point-wisely $\Z_2$ fixed appears
and a pair of fivebranes wrapped on it are created.}
\end{figure}

The same configuration should also arize if we start with
the configuration for the theory with bare mass $m$ of the quarks.
The theory flows in the infra-red to
pure Yang-Mills theory, and the configuration
is again given by (\ref{oddYM}) provided it is now
in the space-time described by
$\tilde{y}\tilde{x}=v^2-m^2$ where a pair of sixbranes are located at
$v=\pm m$.
The massless configuration would arize if we put the
sixbranes on top of each other at $v=0$ and then
separate them in the $x^6$-direction.
The equation describing the curve is still given by (\ref{oddYM})
with $N_f=1$.
In order for the resulting configuration to be the same as the one
obtained by Higgsing the $SO(N_c+1)$ theory,
the fivebrane must also wrap twice on the $\CP^1$ cycle.
Since the curve (\ref{oddYM}) generically
does not intersects the $\Z_2$ invariant
cylinder $v=0$, we can draw the following
general conclusion:
Suppose a sixbrane and its mirror image under the $\Z_2$ action
(\ref{newZ2}) approach each other, colide,
and are separated in the $x^6$-direction.
Then, {\it
a fivebrane wrapping twice on the $\CP^1$ cycle is created}
(see Figure \ref{creation}).
This is in contrast with the situation where the $\Z_2$ action is the
simple sign flip of the coordinates $x^{4,5,7,8,9}$: In this case,
the fivebrane should not be created when the sixbranes are separated,
as we can see by looking at the case of $Sp(N_c)$
or $SO({\rm even})$ gauge groups.

\subsection*{\it $SO({\rm odd})$ with General $N_f$}

In principle, we should be able to obtain the
configuration for higher $N_f$ theory with
$SO(N_c={\rm odd})$ gauge group by Higgsing the one for
$SO(N_c+1)$ theory with $N_f+1$ flavors.
However, since we do not presently know a precise map between the
Higgs vevs and the location of the $\CP^1$ components,
this is not a good way to find it.
Instead, we can use the phenomena of fivebrane creation
which we have found above.
We start with a theory where a bare mass $m$ is given to all of the
quarks. The configuration for this is the fivebrane wrapping the
curve (\ref{oddYM}) in the space-time where
$N_f$ pairs of sixbranes are located at $v=\pm m$ and at
different points in the $x^6$-direction.
The space-time is a $\Z_2$-invariant resolution of
$\tilde{y}\tilde{x}=(v^2-m^2)^{N_f}$ where
the $\Z_2$ action is given by
(\ref{newZ2}), that is, $\tilde{y}\to-\tilde{y}$,
$\tilde{x}\to -\tilde{x}$ and $v\to -v$.
Making each pair close to the $\Z_2$ ``fixed plane'' $v=0$
and separating them in the $x^6$ direction,
there appear $2N_f-1$ $\CP^1$ cycles $C_i$ ($i=1,2,\ldots,2N_f-1$ from
the left) where the $\Z_2$ acts trivially on $C_{2i-1}$ and
by $\pi$-rotation on $C_{2i}$.
According to the phenomena found above,
$\CP^1$ components of the fivebranes wrapping twice on
each of $C_{2i-1}$ are created.
In the end, the fivebrane consists of these $\CP^1$ components
in addition to the main component which is
still described by the equation (\ref{oddYM}).
When $r$ of $\phi_a$'s vanish, the curve (\ref{oddYM}) degenerates and
additional $\CP^1$ components are created.
It is a simple exercise to show that
the configuration where all the components are at $x^{7,8,9}=0$
satisfy the consistency requirement coming from
the flux quantization condition;
The number of fivebranes wrapping
each of the $\Z_2$ fixed cycles $C_{2i-1}$ is even
and the number of fivebranes intersecting
each of them is also even.
The number of $\Z_2$ invariant deformation
of the location of $\CP^1$ components subject to the constraint
to avoid {\it t}-configuration is
$(2r+1)N_f-(2r+1)2r/2$ which agrees with the dimension of the
$r$-th Higgs branch where the theory at the root has
$SO(2r+1)$ unbroken gauge group and $N_f$ massless quark
multiplets.

\section{$M$ Theory Realization of Orientifold Four-Plane}

\subsection{$\rm O4$-plane with a Single ${\rm D}4$-brane
Stuck on it}

From the discussion in the previous
section, we learned of an \MT realization
of $SO$-type O4-plane with a single D4-brane stuck on it.
Note that the na\"\i ve candidate --- \MT on
$\R^5\times\R^5/\Z_2\times S^1$ with a single fivebrane stuck
at the $\Z_2$ fixed point --- has been ruled out
by the flux quantization condition \cite{Wflux} (see section 2).
We propose that the correct realization
is \MT on $\R^5\times(\R^5\times S^1)/\Z_2$
where the $\Z_2$ acts on $\R^5\times S^1$
with coordinates $x^{4,5,7,8,9}$ and $x^{10}$ as (\ref{newZ2}):
\beq
\begin{array}{l}
x^{4,5,7,8,9}\to-x^{4,5,7,8,9}\\[0.1cm]
x^{10}\to x^{10}+\pi.
\end{array}
\label{newZ22}
\eeq
Because this $\Z_2$ action is free, as has been said,
there is no fivebrane charge that would be
associated with the $\R^5/\Z_2$ fixed point \cite{DM,orbW}.
In the weakly coupled Type IIA limit where
the size of $S^1$ shrinks to zero, the $\Z_2$-invariant
cylinder $x^{4,5,7,8,9}=0$ looks like a $\Z_2$
fixed plane in ten-dimensional space-time but it carries no D4-brane
charge.
This is the right property of a D4-brane stuck on the $SO$-type
O4-plane. Further test of this proposal will be discussed
elsewhere.

One can show that \MT on this space satisfies the flux quantization
condition. First of all, as said above, there is nothing
carrying the fivebrane charge and therefore $G=0$.
On the other hand, the space $(\R^5\times S^1)/\Z_2$ can be considered
as the M\"obius bundle over $S^1$ and is homotopy equivalent to
the base $S^1$ as the $\R^5$ fibre can be contracted;
$(\R^5\times S^1)/\Z_2\simeq S^1$. The fourth
Stiefel Whitney class of this space of course vanishes.
Thus, the condition (\ref{Flux}) is satisfied.

As an application, we propose the \MT realization of the
component of the moduli space of
Type IIA orientifold on $T^5/\Z_2$ where
a single D4-brane is at each of the 32 fixed points.
As noted in \cite{orbW}, this is dual to a component of the moduli
space of Type I string theory on $\tilde{T}^5$ (the dual of $T^5$)
which corresponds to a component of the moduli space of flat
${\rm Spin}(32)/\Z_2$ connections on $\tilde{T}^5$ that is
not connected through a family of flat connections to
the component of trivial connection.
We propose that it is dual to
\MT on $\R^5\times (T^5\times S^1)/\Z_2$ where $\Z_2$ acts on
$T^5$ as the sign flip of the five coordinates and on $S^1$ as the
$\pi$-rotation.

\subsection{$\rm O4$-plane of $Sp$-Type}

The discussion in section 5 also suggests a new \MT
realization of $Sp$-type O4-plane.
It is \MT on $\R^5\times(\R^5\times S^1)/\Z_2$
with the $\Z_2$ action (\ref{newZ22})
with a single fivebrane wrapped on
the invariant cylinder at $x^{4,5,7,8,9}=0$.
It has the right D4-brane charge ($+1$) since 
the fivebrane charge is provided only by the
single fivebrane wrapped on the invariant cylinder.
In addition, the wrapped fivebrane cannot be deformed
without an energy of order $1/R$, where $R$ is the radius of
$S^1$ which is related to the Type IIA string coupling
by $R=\gst\elst=\gst^{2/3}\elel$,
and therefore its motion is frozen in the perturbative
string regime $\gst\ll 1$,
as expected since orientifold four-plane has no
dynamical degree of freedom.\footnote{It is interesting
to find the interpretation of the fivebrane with its
$x^{4,5,7,8,9}$-location varying as a function of
$x^{10}$ with wavelengths $\sim R$. A natural candifate is a bound
state of D0-branes with O4-plane. It is intetresting to
confirm the existence of such a bound state using the
matrix quantum mechanics.}

As a consistency check, we examine whether this configuration
 satisfies the flux
quantization condition (\ref{Flux}).
We choose a $\Z_2$ invariant four-cycle $\hS=S^4_0\cup S^4_{\pi}$
in the double cover where $S^4_0$ is a four-sphere at $x^{10}=0$
in the $x^{4,5,7,8,9}$-plane surrounding the origin
and $S^4_{\pi}$ is its $\Z_2$ partner at $x^{10}=\pi$.
We give the natural orientation to both of the spheres so that
the $\Z_2$ action reverses the orientation of $\hS$.
Then, the flux of $G/2\pi$ through $\hS$ is defined to be
2 and hence the flux through the quotient $S=\hS/\Z_2$ is 1.
On the other hand $w_4=0$
for this space as noted above. Therefore, the condition
(\ref{Flux}) is satiafied.

We note that there is
another realization of $Sp$-type O4-plane --- \MT on
$\R^5\times\R^5/\Z_2\times S^1$ with a pair of fivebranes
frozen at the $\Z_2$ fixed plane --- which we have actually used
in the construction of brane configuration
for symplectic gauge theory.
This O4-plane has the right D4-brane charge ($-1+2=+1$), but
the fivebrane pair can na\"\i vely separated from the
$\Z_2$ fixed plane without non-zero energy,
which is not the property of the O4-plane of $Sp$-type.
However, the pair of fivebranes can be made fixed at the
$\Z_2$ fixed plane by imposing a particular boundary condition
so that we can realize the $Sp$-type O4-plane
ending on an NS5-brane, as we have seen in section 3.
Another example of freezing mechanism will be given in the following
section and it also yields a finite (or semi-infinite)
O4-plane of $Sp$-type.
At present, we do not know the freezing mechanism
to realize an infinite O4-plane of $Sp$-type.
This mysterious freezing is presumably related to
the frozen $D_4$ ($D_8$) singularity in $M$ ($F$) theory
that corresponds to orientifold
six-(seven-)plane of $Sp$ type \cite{Wtor}.

Thus, we have two realizations of orientifold four-plane
of $Sp$-type. Are the $Sp(n)$ gauge theories on the worldvolume
of $n$ pairs of D4-branes close to them equivalent?
Here we note that $\pi_4(Sp(n))=\Z_2$ and
that there is a $\Z_2$ valued ``theta angle''
corresponding to assigning
a weight $+1$ or $-1$ to the path-integral on the topologically
non-trivial sector.
It is tempting to suspect that the two choices of
this theta angle corresponds to the two realizations of
$Sp$-type O4-plane.
We will come back to this point in section 6.5.

\subsection{RR $U(1)$ Gauge Field}

\newcommand{\RRo}{A^{\rm RR}_1}
\newcommand{\tilO}{\widetilde{\rm O4}^+}
\newcommand{\bx}{{\bf x}}

When we consider \MT on $(M\times S^1)/\Z_2$
where the generator of $\Z_2$ acts
as an involution $\gamma$ of ten-dimensional space-time
$M$ and as $\pi$-rotation on the circle $S^1$ in the eleventh
direction, there is possibly a``flux'' of the RR one-form
$\RRo$ in the corresponding Type IIA string theory on $M/\Z_2$
\cite{SchS}.
This can be explained as follows.
The wavefunction of a single D0-brane in this situation
is an odd function
of the space-time $M$, $\psi(\gamma \bx)=-\psi(\bx)$,
$\bx\in M$.
In other words, it is a section of the
complex line bundle associated with the $U(1)$-bundle
$(M\times U(1))/\Z_2$ over $M/\Z_2$ where this $U(1)$ group
is identified with the circle $S^1$ in the eleventh direction.
Then,
the RR one-form $\RRo$ represents a connection of this
$U(1)$-bundle since D0-branes are electrically charged under $\RRo$.
It is determined by the metric of the eleven-dimensional space-time
in such a way that the directions orthogonal
to $U(1)$ fibre are horizontal.
When the metric of the eleven-dimensional manifold
$M\times S^1$ is such that $M$ and $S^1$ are orthogonal,
the connection of the $U(1)$ bundle over $M/\Z_2$ is 
the standard one that is induced from the
trivial flat connection over $M$. In such a case,
the holonomy along the loop represented by a path in $M$
connecting a point $\bx$ and its
$\Z_2$ partner $\gamma \bx$ is $-1$.

In the two new realizations of Type IIA
orientifold on $\R^5/\Z_2$ discovered
above --- corresponding to O4-plane of $SO({\rm odd})$-type
and O4-plane of $Sp$-type ---
\MT geometry is $\R^5\times(\R^5\times S^1)/\Z_2$
where $S^1$ is orthogonal to $\R^5\times \R^5$.
Here, we note that the base ten-dimensional manifold $\R^5\times
\R^5/\Z_2$ is singular and we should delete a
neighborhood of the origin of $\R^5/\Z_2$
as the low energy supergravity approximation
breaks down near such a singularity.
Then,
the RR $U(1)$ gauge field $\RRo$
has a non-trivial holonomy as described above.
Namely, the holonomy along the non-trivial loop in the $\Z_2$
quotient is $-1$.
This is in contrast with the realization of
O4-plane of $SO({\rm even})$-type
and the old realization of O4-plane of $Sp$-type.
In these cases, the \MT geometry is $\R^5\times\R^5/\Z_2\times S^1$
where $S^1$ is orthogonal to $\R^5\times\R^5$, and therefore
the RR gauge field is trivial $\RRo=0$.

We could not find an \MT realization of O4-plane of
$SO({\rm odd})$-type with a trivial RR $U(1)$ gauge field.
The only candidate one can think of, \MT on $\R^5\times\R^5/\Z_2\times S^1$
with a single fivebrane at the $\Z_2$ fixed plane,
is forbidden by the quantization condition (\ref{Flux})
(see section 2.1).
This fact strongly suggests that a non-trivial RR Wilson line
is {\it required} for O4-plane of $SO({\rm odd})$-type by some
non-perturbative reason.\footnote{
If the precise duality map between Type IIA orientifold on $T^5/\Z_2$
with one D4-brane at each O4-plane
to Type I string theory with a non-trivial
${\rm Spin}(32)/\Z_2$ Wilson line were available,
one could test this as in \cite{FHSV}.}
Similarly, the fact that we cannot find an \MT realization of
O4-plane of $SO({\rm even})$-type with a non-trivial RR Wilson line
implies that such an O4-plane does not exist.

One thing that is possibly related to this
is the Dirac quantiztion condition for D2-branes in Type IIA string
theory. This may be derived by repeating the derivation
in \cite{Wflux} of (\ref{Flux}).
In the background with flat RR gauge field, the CS coupling of
\cite{inflow} is identical to that of the membrane in $M$ theory,
and hence we will get the same condition
for the field strength of the RR three-form potential.
However, if that is true,
by considering the four-cycle surrounding the $\R^5/\Z_2$ ``fixed
plane'' in ten dimensions, we find a contradiction since $\int w_4$
is $1$ mod 2 as usual while O4-plane of $SO({\rm odd})$-type has
even unit of magnetic charge.
As we will see in a different context, there is a subtlety
in the definition of CS coupling, and this should be the reason for
the apparent contradiction.
In any case, the ten-dimensional flux quantization
condition must be modified so that $SO({\rm odd})$-type O4-plane is
allowed. It is possible that the modified condition requires
non-trivial RR Wilson line for $SO({\rm odd})$ O4-plane,
and trivial one for $SO({\rm even})$ O4-plane.

\medskip
\noindent
{\bf To summarize},
we list up the orientifold four-planes and their \MT realizations.

\noindent
O4${}^-$:~
the O4-plane of $SO$-type (D4-brane charge $-1$) with the
trivial RR $U(1)$ gauge field. This is
realized as \MT on $\R^5\times\R^5/\Z_2\times S^1$.

\noindent
O4${}^0$:~
the O4-plane of $SO$-type with a single D4-brane stuck on
it (D4-brane charge $0$) with a non-trivial
RR $U(1)$ gauge field (holonomy$=-1$ along a non-trivial loop).
This is realized as \MT on $\R^5\times(\R^5\times S^1)/\Z_2$.

\noindent
O4${}^+$:~
the O4-plane of $Sp$-type (D4-brane charge $+1$)
with the trivial RR $U(1)$ gauge field.
This is realized
as \MT on $\R^5\times\R^5/\Z_2\times S^1$ with a pair of fivebranes
frozen at the $\Z_2$-fixed plane.

\noindent
$\tilO$:~
the O4-plane of $Sp$-type (D4-brane charge $+1$)
with a non-trivial RR $U(1)$ gauge field (holonomy$=-1$ along
a non-trivial loop).
This is realized as \MT on $\R^5\times(\R^5\times S^1)/\Z_2$
with a single fivebrane stuck on the $\Z_2$ invariant cylinder.

\subsection{T-duality to Orientifold Five-plane}

There is a similar pattern in the list of orientifold five-planes
in Type IIB string theory.
In \cite{WnewG}, it was found that the O5-planes
of $SO({\rm even})$-type as well as of
$Sp$-type can be realized in a background
with trivial Type IIB theta angle (RR zero-form) $\theta_B=0$,
whereas the O5-planes of $SO({\rm odd})$-type as well as
of $Sp$-type
can (also) be realized in a background with $\theta_B=\pi$.
Here we show that this actually
follows from what we have learned above
via T-duality.
We show in particular that Type IIA orientifold with
the non-trivial RR $U(1)$ gauge field is mapped
to Type IIB orientifold with $\theta_B=\pi$.

We compactify the $x^4$ direction on a circle and consider $x^4$
as a dimensionless periodic coordinate of period $2\pi$.
We consider Type IIA orientifold on $\R^5\times (S^1\times \R^4)/\Z_2$
where $\Z_2$ acts as the sign flip of the five coordinates
$x^{4,5,7,8,9}$.
There are two orientifold four-planes --- one at
$x^4=x^{5,7,8,9}=0$ and the other at $x^4=\pi,x^{5,7,8,9}=0$.

If both of them are O4${}^-$-plane, then by T-duality
we will obtain Type IIB orientifold on
$\R^5\times S^1\times\R^4/\Z_2$ where a single O5-plane of
$SO$-type (D5-brane charge $-2$) wraps on the $\Z_2$ fixed plane
$\R^5\times S^1$.
Similarly, if both are O4${}^+$, we obtain O5-plane of $Sp$-type
(D5-brane charge $+2$) wrapped on $\R^5\times S^1$.
In both of these cases, since RR one-form is zero on the Type IIA
side, RR zero-form $\theta_B$ is zero.

Next let us consider the case where the O4-plane at $x^4=0$
is O4${}^-$ and the one at $x^4=\pi$ is O4${}^0$.
After T-duality, we obatin O5-plane of $SO$-type
with a single D5-brane stuck on it (D5-brane charge $-1$)
wrapped on $\R^5\times S^1$. 
What is the RR $U(1)$ gauge field in such a system and what does it
correspond to in the Type IIB side?
It should be trivial around O4${}^-$
but it should be non-trivial around O4${}^0$
so that the wavefunction satisfies $\psi(\gamma\bx)=-\psi(\bx)$
near $x^4\sim \pi$.
A natural candidate is such that the wavefunction satisfies
$\psi(\gamma\bx)=\e^{ix^4}\psi(\bx)$.
Namely, the $U(1)$ bundle is
$\R^5\times (S^1\times \R^4\times U(1))/\Z_2$
where the $\Z_2$ action is
\beq
(x^{0,1,2,3,6},\e^{ix^4},x^{5,7,8,9},\e^{ix^{10}})
\longrightarrow (x^{0,1,2,3,6},\e^{-ix^4},-x^{5,7,8,9},\e^{i(x^4+x^{10})}),
\eeq
where $x^{10}$ is identified as the coordinate of the group manifold
$U(1)$.
The trivial flat connection on the trivial $U(1)$ bundle
on the double cover $\R^5\times S^1\times \R^4$
does not descend to a connection of this bundle.
However, a certain non-trivial one does.
Let us choose a flat connection on the double cover,
and let $g(\bx)$ be a $U(1)$ valued (local) function of
$\R^5\times S^1\times \R^4$ that determines a horizontal
section $\bx\mapsto(\bx,g(\bx))$
(i.e. the gauge field is given by $iA=g^{-1}\dd g$).
This descends to the $\Z_2$ quotient if and only if
\beq
\e^{ix^4}g(\gamma\bx)\,=\,g(\bx).
\eeq
Up to single valued gauge transformations, there is a unique solution
\beq
g(\bx)=\e^{ix^4/2}.
\eeq
Namely, the RR $U(1)$ gauge field in the double cover
is $\RRo={1\over 2}\dd x^4$. By the standard relation
between RR $p$-form and RR $(p\pm 1)$-form under the T-duality,
we see that the Type IIB theta angle is given by
\beq
\theta_B=\oint\limits_{S^1} \RRo=\pi.
\eeq
The same thing holds if we have one O4${}^+$ and one $\tilO$.
In such a case, we will obtain O5-plane of $Sp$-type
(D5-brane charge $+2$) wrapped on $\R^5\times S^1$
in a background with $\theta_B=\pi$.

Through this argument, we have also given the \MT realization of
O5-plane wrapped on $\R^5\times S^1$.
The eleven-dimensional geometry is nothing but the total space
of the RR $U(1)$ bundle where the metric is determined so that
the $U(1)$ fibre is orthogonal to
the horizontal directions defined by the RR gauge field.
In particular, in the case of non-trivial RR gauge field considered above,
not all of the $x^{0,1,\ldots,9}$
directions at constant $x^{10}$ are orthogonal to
the fibre direction,
but the two-torus in the $x^{4}$-$x^{10}$ directions
is twisted so that
$\delta_v(x^{4},x^{10})=(0,\epsilon)$ is orthogonal to
$\delta_h(x^{4},x^{10})=(\epsilon, \epsilon/2)$.
This is another explanation of $\theta_B=\pi$.

\subsection{The $\Z_2$ Valued Theta Angle}

In this section, we have seen that there are
two kinds of O4-planes and O5-planes of $Sp$ type
which differ in the RR potentials.
In five- and six-dimensional symplectic gauge theories,
there is a $\Z_2$ valued theta angle associated with
$\pi_4(Sp(n))=\Z_2$ and $\pi_5(Sp(n))=\Z_2$.
This tempts us to suspect that the two kinds of O4-planes
or O5-planes correspond to the two choices of the theta angle
(as already mentioned in \cite{WnewG} for O5 case).

To see if that is the case, one needs to know the coupling of
RR potentials and gauge fields on the D-brane worldvolume.
Such a coupling, called Chern-Simons (CS) or Wess-Zumino term,
is extensively studied
and a certain understanding has been obtained
\cite{inflow,CZ,MM}
up to some global issues related to the following discussion.
For $n$ D-branes on top of one another, with worldvolume $W$,
it takes the form $\int_W C{\rm ch}$ or
$\int_W n C-\int_W H {\rm ch}^{(0)}$
where $C$ ($H$) is a sum of RR potentials (field strengths)
and ${\rm ch}$ is the
Chern character of the $U(n)$ bundle on the D-brane
and ${\rm ch}^{(0)}$ is the associated Chern-Simons form
(we ignore
the contribution from the normal and tangent bundle of $W$
and also from the NS two-form).
In particular, for D5-branes, the term $\int\theta_B{\rm ch}_3$
shows that the Type IIB theta angle defines the theta angle of $U(n)$
gauge theory associated with $\pi_5(U(n))=\Z$.

It is natural to expect that the CS coupling takes a similar form
even when these D-branes wrap over an orientifold plane.
\footnote{Of course it is important to determine it.
See \cite{DJM} for some discussion.}
For D5-branes on top of an O5-plane, it may appear that we obtain
the theta term with theta angle $\theta_B$ as in D-brane case.
However, ${\rm ch}_{\rm odd}$ vanishes for
symplectic (orthogonal) bundles since the curvature changes by sign
flip under the transposition,
${}^tF_A=-JF_AJ^{-1}$ (${}^tF_A=-F_A$).
In fact for $Sp$ bundles there is no characteristic {\it cohomology}
class (over an arbitrary coefficient) at $(4k+2)$-dimension
since the classifying space $BSp$ is a quaternionic Grassmannian
which has a cell decomposition into $4k$-dimensional cells.
Therefore, what can distinguish the non-trivial bundle
must be something more global than just an integral of some
cohomology class.

Actually, there is a subtle topological problem in the definition of
CS coupling.
When the RR potentials and the gauge fields on the D-brane
are both topologically
non-trivial, the expression for the CS coupling written above
does not actually make sense.
In three-dimensional gauge theory, there is a standard recipe for 
defining the CS term $\int {\rm ch}_2^{(0)}$
for topologically non-trivial configuration
\cite{DijkW}:
First find a four-manifold bounded by
the three-manifold of interest, 
extend the gauge field to the interior of the four-manifold,
and then evaluate $\int {\rm ch}_2$ on it. In general the extension is
impossible but the difference of the values of the CS term can be
similarly defined. Therefore, one can define the CS term by taking
a reference gauge field and assigning a suitable value of it
for such a reference configuration.
One can find such an assignment
so that the basic physical requirements are satisfied \cite{DijkW}.
Such a consistent assignment is not in general unique
and one must make a choice
among several possibilities, which is analogous to the choice
of theta angle.
What we need here is a generalization of this to the case where
the gauge theory lives on a submanifold of a ten-dimensional
space-time and there are also
RR field strengths in addition to gauge fields.
A systematic study of such a topological coupling has not been done
yet. (There is, however, a related discussion in \cite{Meff}.)
It is possible in the case of O4- or O5-plane of $Sp$ type
that there are two choices of a consistent assignment
depending on the two choices of RR Wilson line or $\theta_B$
and that these corresponds to the $\Z_2$-valued theta angle.

\section{Applications}

In this section, we consider
applications of the consistency condition derived in section 2
and the \MT realization of
orientifold four-planes proposed in section 6.
In particular
we study the properties of orientifold four-planes
intersecting with D6-branes and/or NS5-branes.

\subsection{O4-D6 System}

We first analyze the system of an O4-plane intersecting with
D6-branes.
Let us consider a Type IIA orientifold
by the sign flip of the coordinates $x^{4,5,7,8,9}$ 
when there is a D6-brane at $x^{4,5,6}=0$ spanning the
$x^{0,1,2,3,7,8,9}$-directions.
Then, the orientifold plane at $x^{4,5,7,8,9}=0$ intersects
with the D6-brane at $x^6=0$ and is divided into two
parts --- the left part $x^6<0$ and the right part $x^6>0$.
We determine what types of O4-plane is possible for
the two parts.

A single D6-brane at $x^{4,5,6}=0$ is realized in \MT
as the Taub-NUT space in the $x^{4,5,6,10}$-direction.
With a choice of the complex structure
such that $v=x^4+ix^5$ is a complex coordinate,
the Taub-NUT space is described by
\beq
yx=v,
\label{A0}
\eeq
where $y$ and $x$ are other complex coordinates
expressed as $y=\e^{-x^6/R-ix^{10}}f(x^{4,5,6})$
and $x=\e^{x^6/R+ix^{10}}g(x^{4,5,6})$.
Here $R$ is the radius at $x^6\to\pm\infty$ of the circle
in the eleventh direction $x^{10}$, and
$f(x^{4,5,6})$ and $g(x^{4,5,6})$ are some functions 
(the precise form of $f$ and $g$
is necessary only in the next subsection and
will be given there).
As a complex manifold, it is the same as the complex plane $\C^2$
with coordinats $(y,x)$.
The locus $x^{4,5}=0$ consists of two components corresponding
to the left $x^6<0$ and the right $x^6>0$ parts:
The left part is the $y$-axis $\{x=0\}$ and the right part is
the $x$-axis $\{y=0\}$.

We consider $\Z_2$ orbifold of \MT on this space by
the action
\beq
y\to y,~~x\to -x,~~ v\to -v~~~\mbox{and}~~
x^{7,8,9}\to -x^{7,8,9}.
\label{ZZ}
\eeq
We can consider this as an \MT realization of an
O4-plane intersecting with the D6-brane
since the $\Z_2$ action
reduces to the sign flip of $x^{4,5,7,8,9}$
in the weakly coupled Type IIA
string theory limit $R\to 0$.
(\ref{ZZ}) is essentially the unique one that reduces
to such a $\Z_2$ action
(there is actually another one, but it is simply related by
the interchange of $x$ and $y$ (i.e. sign flip of $x^6$)).
On the left part $x^6<0$, the $\Z_2$ acts as the sign flip of
the five coordinates fixing the eleventh coordinates,
while it acts as the sign flip of five coordinates
together with the $\pi$-shift of the eleventh coordinate
on the right part $x^6>0$.
Therefore, the O4-plane on the left part is
O4${}^-$ (the O4-plane of $SO$-type, D4-brane charge $-1$,
trivial RR $U(1)$ holonomy),
and the O4-plane on the right part is
O4${}^0$ (the O4-plane of $SO$-type with a single D4-brane stuck on it,
no D4-brane charge, non-trivial RR $U(1)$ holonomy).
This is one possible pattern of dividing O4-plane by D6-brane.
It is free to wrap the fivebrane even number of times
on each of the $y$-axis and the
$x$-axis.
This corresponds to putting even number of D4-branes
on each of the two parts.
Therefore, it is possible to have O4${}^-$ plus $2n$ D4-branes
on one side and  O4${}^0$ plus $2m$ D4-branes on the other.

By the flux quantization condition, it is not possible to
wrap odd number of fivebranes on the $y$-axis
since $y$-axis is the $\Z_2$ fixed plane.
On the other hand, it is possible to wrap odd number of
fivebranes on the $x$-axis provided
the $y$-axis at $x^{7,8,9}=0$ (the $\Z_2$ fixed plane)
is screened by at least one pair of fivebranes.
We consider the minimal case where
the fivebrane wraps twice on the $y$-axis and once on the $x$-axis.
The fivebrane wrapped on the $x$-axis cannot move in the
$x^{7,8,9}$-directions by the $\Z_2$-invariance.
The pair of fivebranes wrapped on the $y$-axis are also frozen;
if they were separated in the $x^{7,8,9}$-directions
a {\it t}-configuration would appear.
Thus, the left and the right parts both correspond to O4-plane of
$Sp$-type, but one is O4${}^+$ (trivial RR $U(1)$ holonomy)
and the other is $\tilO$ (non-trivial RR $U(1)$ holonomy).
This is another
pattern of dividing O4-plane by D6-brane.
It is possible to add even number of fivebranes on both sides.
Namely, it is also possible to have O4${}^+$ plus $2n$ D4-branes
on one side and  $\tilO$ plus $2m$ D4-branes on the other.

\begin{figure}[htb]
\begin{center}
\epsfxsize=4in\leavevmode\epsfbox{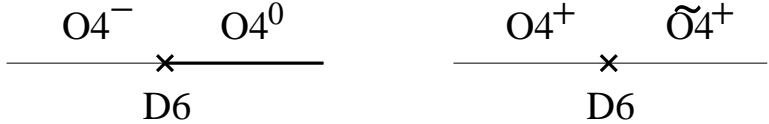}
\end{center}
\caption{D6-brane Dividing O4-plane}
\label{Break}
\end{figure}

Are these the only possibilities?
By the action of $\Z_2$ (\ref{A0}) on the Taub-NUT space,
RR $U(1)$ gauge field must be trivial on one side
and non-trivial on the other. This corresponds to the fact that
the D6-brane is a magnetic monopole for the RR $U(1)$ gauge field
$\RRo$,
and the side with non-trivial holonomy corresponds to the Dirac string
in the following sence.
Let us consider a small circle in the $x^{4,5,6}$-space
encircling the O4-plane in one side (say $x^6<0$)
and move it to the other side ($x^6>0$).
Then the circle sweeps out a cylinder that
is a part of a two-sphere surrounding the D6-brane at $x^{4,5,6}=0$.
If the integral of $\RRo$ along the circle
is zero at the starting point, it increases as the cicrle moves from
$x^6<0$ to $x^6>0$ and finaly becomes $2\pi$ since the D6-brane 
carries a unit magnetic charge.
Therefore, the integral along the half-circle
(which is a closed circle in the $\Z_2$ quotient)
changes from $0$ to $\pi$ as it moves from one side to the other.

This exclude many possibilities such as O4${}^-$-D6-O4${}^+$.
The remaining cases to consider are
O4${}^-$-D6-$\tilO$, and
O4${}^+$-D6-O4${}^0$.
In the first case, a single fivebrane wraps on the $x$-axis
and it intersects transversely with the bare $\Z_2$ fixed plane,
which is impossible by the consistency condition.
The second case may apparently be realized by a pair of fivebranes
screening the $y$-axis at $x^{7,8,9}=0$. However, since there is nothing
wrapped on the $x$-axis, the pair of fivebranes can freely
be separated in the $x^{7,8,9}$-directions. Therefore it is more
natural to consider the configuration as O4${}^-$ plus two D4-branes
on the left side and O4${}^0$ on the right side.
This is also what is expected:
$SO$ and $Sp$ type O4-planes are distinguished by sign of the $\RP^2$
diagram of the fundamental string,
but it remains the same as $\RP^2$ moves from one side
to the other since D6-brane has no charge under
the NS two-form potential $B^{\rm NS}_2$.

Thus, we conclude that only
O4${}^-$-D6-O4${}^0$ and O4${}^+$-D6-$\tilO$ as depicted in Figure
\ref{Break}
(plus even number of D4-branes on both sides)
are the allowed patterns of dividing an O4-plane
by a single D6-brane.

\subsection*{\sl O5-D7 System}

Here we digress for mentioning the implication of this
on the system of O5-plane and D7-branes in Type IIB string theory.
We compactify the $x^4$-direction as in section 6.4
in the presence of a single D6-brane at $x^{4,5,6}=0$.
The $\Z_2$ fixed plane at $x^4=0$ is divided into two parts
by the D6-brane while the one at $x^4=\pi$ is not.
Let us place O4${}^-$ for the part $x^6<0$ of the fixed plane
at $x^4=0$ and O4${}^0$ for the other part $x^6>0$.
We put O4${}^-$ at the $x^4=\pi$ fixed plane.
After T-duality, the D6-brane becomes D7-brane wrapped on the
dual torus, and the $\Z_2$ acts on the space-time
as the sign flip of the coordinates $x^{5,7,8,9}$.
The part with $x^6<0$ of the $\Z_2$ fixed plane $x^{5,7,8,9}=0$
becomes O5-plane of $SO$-type whereas the part $x^6>0$
becomes O5-plane of $SO$-type with a single D5-brane stuck
on it. From what we have learned in section 6.4,
we see that the value of the RR scalar $\theta_B$ is zero on the
$x^6<0$ side while it is $\theta_B=\pi$ on the other side $x^6>0$.
This is always true as far as we start with a single D6-brane,
no matter which type of O4-planes we choose.
That the value of $\theta_B$ varies around the D7-brane is
a familiar fact \cite{Fth} which follows from
the fact that the D7-brane is magnetically charged under
the RR scalar.
Since the $\Z_2$ flips the sign of $\theta_B$,
only $\theta_B=0$ or
$\theta_B=\pi$ are allowed on the orientifold fixed plane
and both in fact appears at the same time.

\subsection*{\sl Splitting D6-branes in O4-plane}

It is interesting to consider the
following question in Type IIA string theory.
Suppose there is an O4-plane of some kind at
$x^{4,5,7,8,9}=0$ and a pair of D6-branes at
$v:=x^4+ix^5=\pm m$, $x^6=0$. The question is what happens when
we make the D6-branes approach each other $m\to 0$,
and then separate in the $x^6$ direction.
\begin{figure}[htb]
\begin{center}
\epsfxsize=4.8in\leavevmode\epsfbox{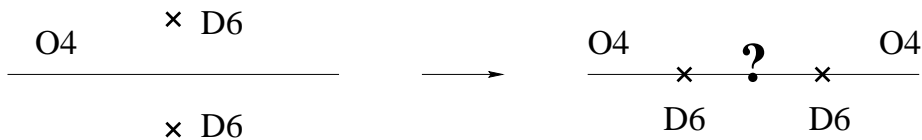}
\end{center}
\caption{The Question}
\label{Q}
\end{figure}
The answer would of course depend on the type of the starting
O4-plane.
We consider the four basic cases: O4${}^-$, O4${}^0$, O4${}^+$
and $\tilO$.

We begin with the case where the starting O4-plane is O4${}^-$.
The starting configuration is realized as
\MT on orbifold $\R^4\times (K\times \R^3)/\Z_2$ where
$K$ is the Taub-NUT space
described by $yx=v^2-m^2$ and the $\Z_2$ action is given by
$(y,x,v,x^{7,8,9})\to (y,x,-v,x^{7,8,9})$.
If we send $m\to 0$, the Taub-NUT space $K$ develops
the $A_1$ singularity $yx=v^2$. Splitting the sixbranes in the $x^6$
direction corresponds to resolving the singularity.
This is described by two coordinate systems $(y_1,x_1)$
and $(y_2,x_2)$ (see section 4) which are related to each other
and to $y,x,v$ by $y=y_1=y_2^2x_2$, $x=y_1x_1^2=x_2$, $v=y_1x_1
=y_2x_2$, $x_1y_2=1$. There is a $\CP^1$ cycle defined by
$y_1=x_2=0$. The $\Z_2$ action is given by
$(y_1,x_1)\to(y_1,-x_1)$ and $(y_2,x_2)\to (-y_2,x_2)$.
Thus, the $\Z_2$ group acts on the $\CP^1$ cycle as
the $\pi$-rotation around
the left D6-brane (at $y_1=x_1=0$) and the right D6-brane
(at $y_2=x_2=0$).
According to the proposal,
the region sandwiched between the two
D6-branes corresponds to O4${}^0$-plane in the weakly coupled
Type IIA limit.

We next consider the case where the starting O4-plane is O4${}^0$.
The starting configuration is \MT on
$\R^4\times (K\times \R^3)/\Z_2$
where the $\Z_2$ action on $K$ is now $(y,x,v)\to (-y,-x,-v)$.
After splitting the D6-branes in the $x^6$
direction, $K$ becomes the resolved $A_1$ Taub-NUT space
described in the same way as above, but the
$\Z_2$ acts on the coordinates as
$(y_1,x_1)\to (-y_1,x_1)$ and $(y_2,x_2)\to (y_2,-x_2)$.
In particular, the $\CP^1$ cycle is point-wisely $\Z_2$ fixed.
This is the situation which we have encountered in the
brane realization of $SO({\rm odd})$ gauge theories.
What happened there is that a pair of fivebranes wrapped on the
$\CP^1$ cycle are created.
In the weakly coupled Type IIA limit, the region stretched
between the D6-branes simply becomes the
O4${}^-$-plane with two D4-branes.

We turn to the case where the starting O4-plane
is $\tilO$.
The \MT geometry is the same as in the O4${}^0$ case.
The starting configuration in the present case includes
a single fivebrane wrapped on the
$\Z_2$ invariant cylinder at $v=0, x^{7,8,9}=0$.
After splitting the D6-branes in the $x^6$-direction,
the fivebrane splits to two components
wrapping the two semi-infinite cigars.
These intersect transversely with the $\Z_2$ fixed plane
(= the $\CP^1$ cycle at $x^{7,8,9}=0$) at different points.
In order to avoid a {\it t}-configuration, one pair of
fivebranes must wrap on the $\CP^1$ cycle at $x^{7,8,9}=0$
to screen the $\Z_2$ fixed plane.
A pair of fivebranes stuck by some force at the $\Z_2$ fixed plane
is nothing but the O4${}^+$-plane.

We finally consider the case where the starting O4-plane is
O4${}^+$.
Since we do not currently know the realization of infinite O4${}^+$,
we use the finite one. In particular, we use the finite O4${}^+$
ending on two D6-branes
which appeared in the final configuration of the previous case.
Sending the D6-branes far apart in the $x^6$-directions
we obatin an almost infinite but actually finite
O4${}^+$-plane.
Now let us bring a new pair of D6-branes near the center of this
long but finite O4${}^+$-plane, let them approach each other,
and split them in the $x^6$-directions.
In \MT geometry, the long $\CP^1$ cycle splits to two disjoint
$\CP^1$'s (let us call $C_L$ and $C_R$)
and a new $\CP^1$ in the middle intersecting with
$C_L$ and $C_R$ appears (call it $C_M$).
The $\Z_2$ acts trivially on $C_L$ and $C_R$, but as $\pi$-rotation
on $C_M$ as well as on the two semi-infinite cigars (one on the
left of $C_L$ one on the right of $C_R$).
The fivebrane wraps twice on each of $C_L$ and $C_R$
and once on each of the semi-infinite cigars.
The flux quantization condition is satisfied near $C_L$ and $C_R$
if a single fivebrane wraps on the $\CP^1$ cycle $C_M$ in the
middle.
This $C_M$, on which $\Z_2$ acts as $\pi$-rotation,
corresponds to $\tilO$ in the weakly coupled Type IIA limit.

\begin{figure}[htb]
\begin{center}
\epsfxsize=4in\leavevmode\epsfbox{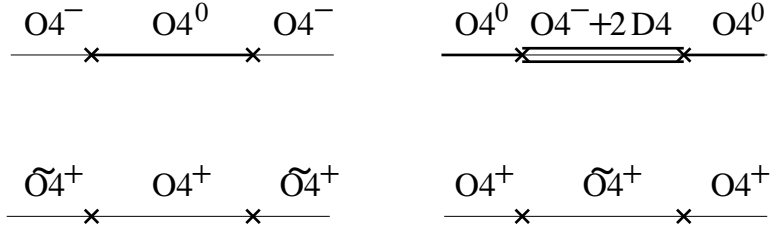}
\end{center}
\caption{The Answer}
\label{A}
\end{figure}
To summarize, we show in Figure \ref{A} the answer to the question.
We see that the answer for
O4${}^0$ case is simply the answer for O4${}^-$ 
plus a single D4-brane.
Also, the answer for $Sp$ type O4-plane is again $Sp$-type O4-plane
irrespective of the type of it.

\subsection{O4-NS5 System Revisited}

Let us next study the system of an O4-plane intersecting with a single
NS5-brane. As D6-brane, NS5-brane divides the O4-plane into two parts.
We are interested in what combination of the types of O4-plane
is possible.

First of all, since NS5-brane does not carry electric nor magnetic
charge for the RR one-form, the RR Wilson line should be
the same on the two parts. This exclude some possibilities
such as O4${}^-\!$-NS-O4${}^0$.

Next, it is not possible to have the same type of O4-plane on
the two sides. This follows from the fact that the NS5-brane has
a unit magnetic charge under the NS two-form potential $B^{\rm NS}_2$.
The sign of $\RP^2$ diagram of fundamental string flips
as $\RP^2$ moves from one side to the other (this argument was first
given in \cite{EGKT}).
This can also be shown by using the consistency
condition for fivebrane and \MT realization of O4-planes
as follows.\\
$\bullet$~ In the case of O4${}^0$ and $\tilO$, it follows trivially from
their \MT realization. They are realized in the eleven-dimensional
space-time
$\R^5\times(\R^5\times S^1)/\Z_2$ where $Z_2$ acts as the sign flip of
$\R^5$ and as $\pi$-rotation of $S^1$.
As usual, we coordinatize the first $\R^5$ factor by $x^{0,1,2,3,6}$
the second $\R^5$ (on which $\Z_2$ acts) by $x^{4,5,7,8,9}$
and $S^1$ by $x^{10}\equiv x^{10}+2\pi$.
If we want a single NS5-brane at $x^{6,7,8,9}=0$ spanning
the worldvolume in the $x^{0,1,2,3,4,5}$-directions,
we must put a single \MT fivebrane at $x^{6,7,8,9}=0$
and at a single point in the $S^1$-direction. This is impossible
since the $\Z_2$ acts freely on $S^1$ and a single fivebrane
at $x^{10}=x^{10}_*$ is accompanied by its mirror
image at $x^{10}=x^{10}_*+\pi$ and the fivebrane
always doubles.\\
$\bullet$~ In the case of O4${}^-$, it is realized by \MT on $\R^5\times
\R^5/\Z_2\times S^1$ and the
 intersection with the NS5-brane is forbidden in order to avoid
a {\it t}-configuration as seen in section 3.\\
$\bullet$~ Finally let us consider the case of O4${}^+$.
Since we do not know the realization
of an infinite O4${}^+$, we consider the problem for the finite
one. Namely, we consider a single NS5-brane intersecting with the
O4${}^+$-plane in the bottom-left configuration of Figure \ref{A}.
In the corresponding \MT configuration, the NS5-brane
is realized as a single fivebrane intersecting transversely
with the $\Z_2$ fixed $\CP^1$-cycle.
However, there are already two fivebranes intersecting with this
$\CP^1$ and therefore it causes an inconsistency;
odd number of fivebranes intersecting with a compact $\Z_2$ fixed
plane.

The remaining possibilities are
O4${}^-$-NS-O4${}^+$ and
O4${}^0$-NS-$\tilO$,
and these are actually possible as we now construct.
Let us introduce complex coordinates $t=\e^{-x^6/R-ix^{10}}$
and $v=x^4+ix^5$ of the $x^{4,5,6,10}$ part
of the space-time $\R^5\times\R^5\times S^1$
where $R$ is the radius of the circle $S^1$ in the
eleventh direction.
Actually, we have already constructed O4${}^-$-NS-O4${}^+$
in section 3. It is given by the weakly coupled Type IIA limit
$R\ll \elel$ of
the fivebrane wrapped on the holomorphic curve
\beq
tv^2=\elst^2,~~x^{7,8,9}=0,
\eeq
in a space-time $\R^5\times \R^5/\Z_2\times S^1$
where we set the unit of length in the
$x^{4,5}$ directions by the string length $\elst$.
 Indeed, as $R\ll \elel$ the fivebrane in the region
$x^6< -\elst$ goes away to infinity $|v|\gg \elst$,
while in the region $x^6>\elst$ it shrinks $|v|\ll \elst$ and
wraps twice on the $\Z_2$ fixed plane.
The $x^6<-\elst$ part of the $\Z_2$ fixed plane is identified
as the O4${}^-$-plane
and the other part $x^6>\elst$
is naturally identified with O4${}^+$-plane
since the fivebrane wraps twice and
it cannot be deformed away from the fixed plane
without cost of energy.
The other one, O4${}^0$-NS-$\tilO$, can be realized as the
$R\ll\elel$ limit of the fivebrane wrapped on the holomorphic curve
\beq
tv=\elst,~~x^{7,8,9}=0,
\eeq
in a space-time $\R^5\times(\R^5\times S^1)/\Z_2$.
(Note that this is invariant under $v\to -v$,
$t\to -t$ ($x^{10}\to x^{10}+\pi$).)
As we take the limit $R\ll\elel$, the fivebrane in the region
$x^6<-\elst$ goes away to infinity $|v|\gg\elst$,
while in the region $x^6>\elst$ it shrinks $|v|\ll\elst$ and
wraps once on the $\Z_2$-invariant cylinder on which
$\Z_2$ acts as $\pi$-rotation.
Thus, the $x^6<-\elst$ part of $\Z_2$ fixed plane is identified
as the O4${}^0$-plane
and the other part $x^6>\elst$ is identified as the $\tilO$-plane.

\subsection{Brane Creation}

It is interesting to see what happens when an NS5-brane passes
through a D6-brane in the presence of an O4-plane.

In the absence of an O4-plane, a D4-brane stretched between them is
created after such a process \cite{HW}.
In addition to the charge conservation
\cite{HW}, there are several arguments to support this.
One direct argument is to use the \MT realization
of Type IIA branes \cite{NOYY,Y}.
We apply it to the case with O4-plane by using its \MT
realization which we have studied above.

Let us consider the Taub-NUT geometry corresponding to a single
D6-brane at $x^{4,5,6}=0$.
As mentioned in the previous subsection,
the complex structure of this space is described
by $yx=v$.
The complex coordinates $y$, $x$ and $v$,
are related to the real coordinates
$x^{4,5,6,10}$ by $v=x^4+ix^5$ and
\beqa
y=\e^{-x^6/R-ix^{10}}\sqrt{\sqrt{|v|^2+(x^6)^2}-x^6},\\
x=\e^{x^6/R+ix^{10}}\sqrt{\sqrt{|v|^2+(x^6)^2}+x^6}\,{v\over|v|}.
\eeqa
Let us consider the fivebrane wrapped on
the curve parametrized by $\zeta$
\beq
y^{n+1}x^n=\zeta.
\label{fami}
\eeq
The equation
is equivalent with $yv^n=\zeta$ and also with $v^{n+1}=\zeta x$
when $x\ne 0$. We note also that
the equation (\ref{fami}) implies
\beq
|v|^n\sqrt{\sqrt{|v|^2+(x^6)^2}-x^6}=\zeta\e^{x^6/R}.
\label{realfa}
\eeq
We will see how the fivebrane configuration looks like
in the limit $R\ll\elel$.

We first consider the case without O4-plane.

\noindent
(i)~
We first take the limit
\beq
\zeta\to \infty,~~~R\ll\elel,~~~\mbox{holding}~~\e^{-L/R}\zeta=1,
\label{1stlim}
\eeq
for some fixed positive $L\gg \elst$.
The right hand side of (\ref{realfa}) is $\e^{(x^6+L)/R}$
and it diverges for $x^6>-L$
but decays to zero for $x^6<-L$ (here and in what follows,
we take $L$ as the
unit of length in the $x^6$-direction and neglect the
error of order $\elst$).
Therefore, after the limit
the fivebrane is at $v=0$ for $x^6<-L$ but blows up to $v=\infty$
at $x^6=-L$. From the equation $yv^n=\zeta$, we see that the
fivebrane wraps $n$-times on the region $x^6<-L$
of the $y$-axis ($v=0, x^6<0$).
This corresponds to a configuration of D6-brane at $x^{4,5,6}=0$
spanning $x^{0,1,2,3,7,8,9}$, NS5-brane at $x^6=-L,x^{7,8,9}=0$
spanning $x^{0,1,2,3,4,5}$ and $n$ D4-branes at $x^{4,5,7,8,9}=0$,
spanning $x^{0,1,2,3}$ and $x^6<-L$.
Namely the NS5-brane is on the left of the D6-brane
and $n$ D4-branes are ending on the NS5-brane from the left.

\noindent
(ii)~
We next take the limit
\beq
\zeta\to 0,~~~R\ll\elel,~~~\mbox{holding}~~\e^{L/R}\zeta=1,
\label{2ndlim}
\eeq
for some fixed positive $L\gg\elst$. 
The right hand side of (\ref{realfa}) is $\e^{(x^6-L)/R}$
and it diverges for $x^6>L$
but decays to zero for $x^6<L$. Therefore, after the limit
the fivebrane is at $v=0$ for $x^6<L$ but blows up to $v=\infty$
at $x^6=L$. From the equation (\ref{fami}), we see that the
fivebrane wraps $n$-times on the $y$-axis
and $(n+1)$-times on the region $0<x^6<L$ of the
$x$-axis ($v=0,x^6>0$).
This corresponds to a configuration of D6-brane at $x^{4,5,6}=0$,
NS5-brane at $x^6=L,x^{7,8,9}=0$
and several D4-branes at $x^{4,5,7,8,9}=0$;
there are $n$ D4-branes spanning $x^{0,1,2,3}$ and $x^6<0$
and $(n+1)$ D4-branes spanning $x^{0,1,2,3}$ and $0<x^6<L$.
Namely, NS5-brane has moved to the right of the D6-brane,
and there are $n$ D4-branes on the left of the D6-brane
as well as $(n+1)$ D4-branes between the D6 and NS5-branes.
We see that a single D4-brane is created when the NS5-brane moves from
$x^6=-L$ to $x^6=L$ passing through the D6-brane at $x^6=0$.

We next include the O4-plane.

\noindent
(1)~
We consider the orbifold by the $\Z_2$ action
\beq
y\to -y,~x\to x,~v\to -v, ~\mbox{and}~x^{7,8,9}\to -x^{7,8,9}.
\label{1stZ2}
\eeq
Then, the $\Z_2$ fixes the $x$-axis at $x^{7,8,9}=0$
point-wisely but acts on the $y$-axis at $x^{7,8,9}=0$
by $\pi$-rotation
of the circle in the eleventh direction.
The equation (\ref{fami}) is $\Z_2$-invariant only if $n$ is odd.
We only consider $n=1$ case which is actually the essential case.

\noindent
(1-ii)~
We first consider the limit (\ref{2ndlim}).
After the limit,
the fivebrane wraps once on the $y$-axis,
wraps twice on the region $0<x^6<L$
of the $x$-axis, and then blows up
at $x^6=L$.
Thus, we have NS5-brane at $x^6=L$ on the right of the D6-brane at
$x^6=0$.
According to our identification of O4-plane,
we have $\tilO$ for $x^6<0$ (on the left of the D6-brane),
O4${}^+$ for $0<x^6<L$ (between D6 and NS5-branes), and
O4${}^-$ for $x^6>0$ (on the right of NS5-brane).\\
(1-i)~
We next consider the limit (\ref{1stlim}).
After the limit,
the fivebrane wraps once on the region $x^6<-L$
of the $y$-axis and then blows up
at $x^6=-L$. Thus, the NS5-brane has moved to $x^6=-L$.
According to the identification,
we have $\tilO$ for $x^6<-L$,
O4${}^0$ for $-L<x^6<0$ and O4${}^-$ for $0<x^6$.
Namely, as the NS5-brane passes through the D6-brane,
the O4${}^+$-plane between them has turned
into O4${}^0$-plane --- O4${}^-$-plane with a single
D4-brane stuck on it.

We next consider the $\Z_2$ action
\beq
y\to y,~x\to -x,~v\to -v, ~\mbox{and}~x^{7,8,9}\to -x^{7,8,9}.
\label{2ndZ2}
\eeq
Then, the $\Z_2$ fixes the $y$-axis at $x^{7,8,9}=0$
point-wisely but acts on the $x$-axis at $x^{7,8,9}=0$
by $\pi$-rotation of the circle in the elventh direction.
The equation (\ref{fami}) is $\Z_2$-invariant only if $n$ is even.
We consider (2) $n=0$ and (3) $n=2$ cases.

\noindent
(2)~
For $n=0$, the fivebrane intersects transversely with the
$\Z_2$-fixed plane (the $y$-axis)
since the equation (\ref{fami}) is $y=\zeta$.
Thus, it is a {\it t}-configuration by itself.
In order to avoid the inconsistency,
we introduce another fivebrane wrapped
twice on the $y$-axis.

\noindent
(2-ii)~
In the limit $\zeta\to 0$ (\ref{2ndlim}),
the fivebrane wraps twice on the $y$-axis
and once on the region $0<x^6<L$
of the $x$-axis, and then blows up
at $x^6=L$.
Thus, we have an NS5-brane at $x^6=L$ on the right of the D6-brane at
$x^6=0$.
The O4-plane is
O4${}^+$ for $x^6<0$, $\tilO$ for $0<x^6<L$, and
O4${}^0$ for $x^6>L$.\\
(2-i)~
In the limit $\zeta\to\infty$ (\ref{1stlim}),
the fivebrane wraps twice on the region $x^6<-L$
of the $y$-axis and then blows up
at $x^6=-L$.
The NS5-brane has moved to $x^6=-L$
and we have O4${}^+$ for $x^6<-L$,
O4${}^-$ with two D4-brane for $-L<x^6<0$
and O4${}^0$ for $0<x^6$.
Namely, as the NS5-brane passes through the D6-brane,
the $\tilO$-plane between them has turned
into O4${}^-$-plane and two D4-branes on top of it are created.

\medskip
\noindent
(3)~
We consider $n=2$ case. In this case
we do not have to introduce
extra fivebranes; the fivebrane wrapping
the curve (\ref{fami}), $y^3x^2=\zeta$, is consistent.

\noindent
(3-i)~
In the limit $\zeta\to\infty$ (\ref{1stlim}),
the fivebrane wraps twice on the region $x^6<-L$
of the $y$-axis and then blows up
at $x^6=-L$.
Thus, we have an NS5-brane at $x^6=-L$
on the left of the D6-brane.
The O4-plane is O4${}^+$ for $x^6<-L$,
O4${}^-$ for $-L<x^6<0$
and O4${}^0$ for $0<x^6$.\\
(3-ii)~
In the limit $\zeta\to 0$ (\ref{2ndlim}),
the fivebrane wraps twice on the $y$-axis
and three-times on the region $0<x^6<L$
of the $x$-axis, and then blows up
at $x^6=L$.
The NS5-brane has moved to $x^6=L$ and
we have O4${}^+$-plane for $x^6<0$ and
$\tilO$-plane with two D4-branes for $0<x^6<L$, and
O4${}^0$-plane for $x^6>L$.
Namely, as the NS5-brane passes through the D6-brane,
the O4${}^-$-plane between them has turned
into O4${}^+$-plane and two D4-branes on top of it are created.

\begin{figure}[htb]
\begin{center}
\epsfxsize=3.6in\leavevmode\epsfbox{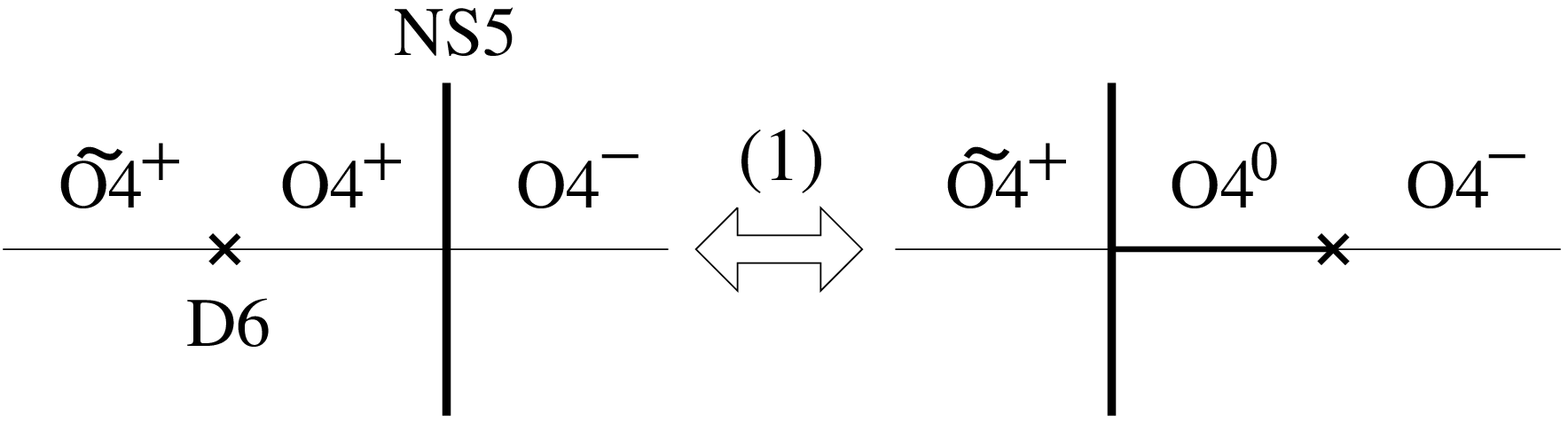}\\[0.1cm]
\epsfxsize=3.6in\leavevmode\epsfbox{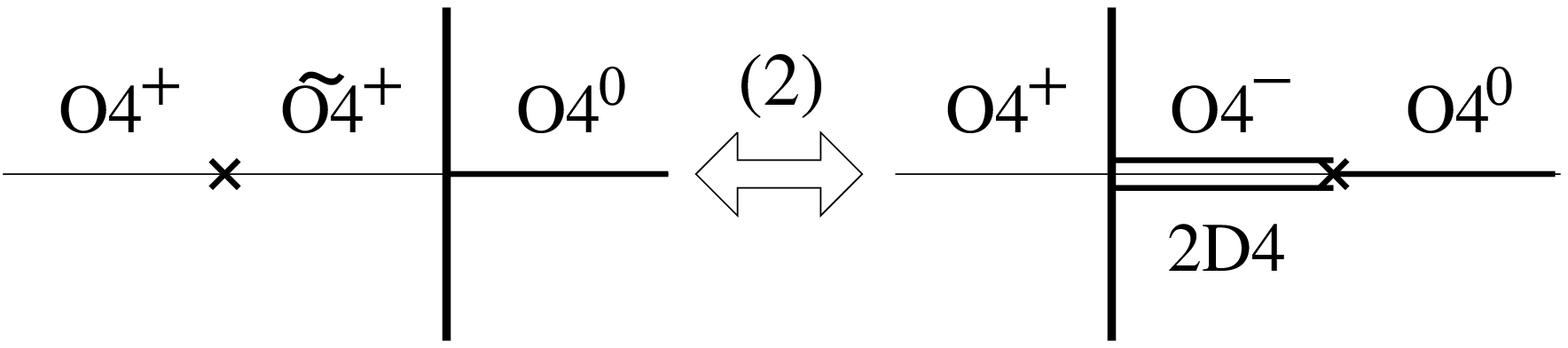}\\[0.1cm]
\epsfxsize=3.6in\leavevmode\epsfbox{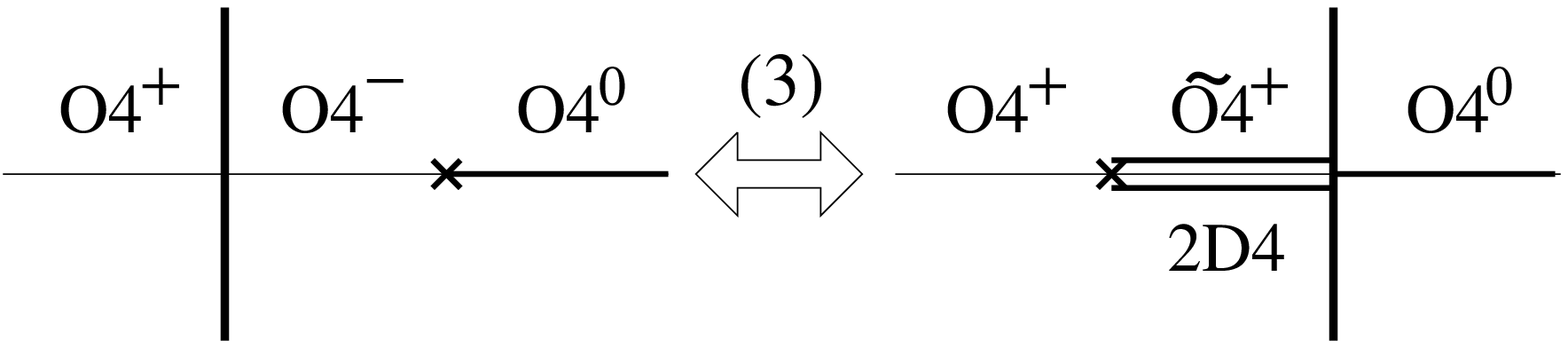}
\end{center}
\caption{The Brane Creation}
\label{Crefig}
\end{figure}
We summarize what we have learned in Figure \ref{Crefig}.
The process (2) is nothing
but the process (1) plus a single D4-brane ending on the NS5-brane
from the right if we neglect the difference in RR $U(1)$ gauge field.

\subsection{s-rule}

It is easy to generalize the {\it s-rule}
to the case with orientifold four-plane.

Suppose there is a D6-brane spanning $x^{0,1,2,3,7,8,9}$
at $x^6=0$ as in the previous subsection
and an NS5-brane spanning $x^{0,1,2,3,4,5}$ at $x^6=L$.
We call it an s-configuration when there are more than one
D4-branes spanning $x^{0,1,2,3}$ and $0<x^6<L$
stretched between D6 and NS5-branes.
S-rule says that an s-configuration is not supersymmetric.
This can be shown by using the \MT realization 
of Type IIA branes as follows.
We use the same notation as in the previous subsection
for the Taub-NUT space describing the D6-brane at $x^6=0$.
$m$ D4-branes ending on the NS5-brane from the left
can be obtained from the fivebrane wrapping the curve
\beq
v^m=\zeta x,
\label{sco}
\eeq
by taking the limit (\ref{2ndlim}).
This equation, when extended to $x^6<0$,
is equivalent to $yv^{m-1}=\zeta$. Therefore, in the
limit (\ref{2ndlim}) the fivebrane also wrapps
$(m-1)$-times on the $y$-axis ($v=0,x^6<0$).
Namely, there are automatically $(m-1)$ D4-branes
on the left of the D6-brane.
Therefore, if there is no D4-branes on the left of the D6-brane,
$m$ must be one (or zero).

\begin{figure}[htb]
\begin{center}
\epsfxsize=1.7in\leavevmode\epsfbox{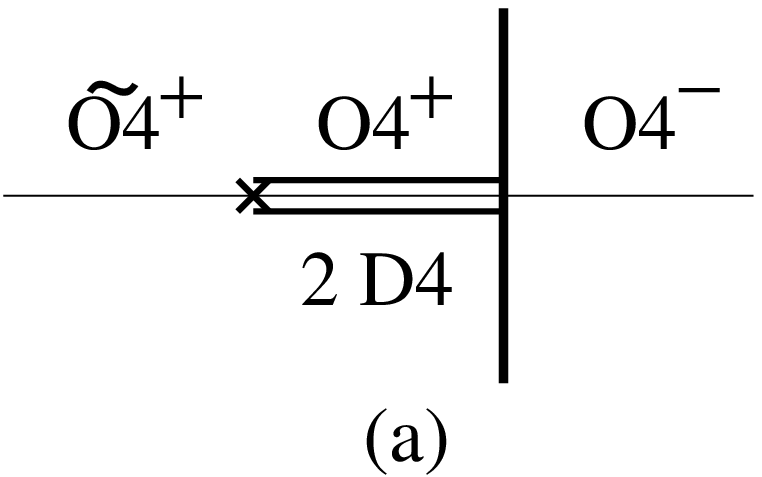}~~~~~
\epsfxsize=1.7in\leavevmode\epsfbox{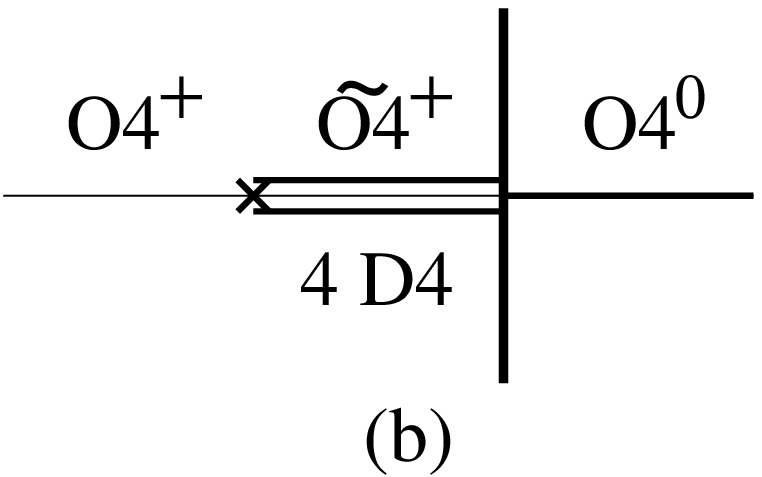}\\[0.5cm]
\epsfxsize=1.7in\leavevmode\epsfbox{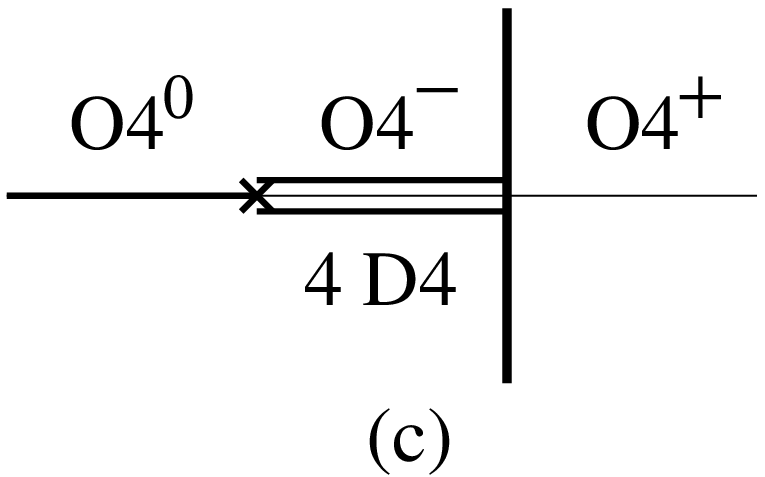}~~~~~
\epsfxsize=1.7in\leavevmode\epsfbox{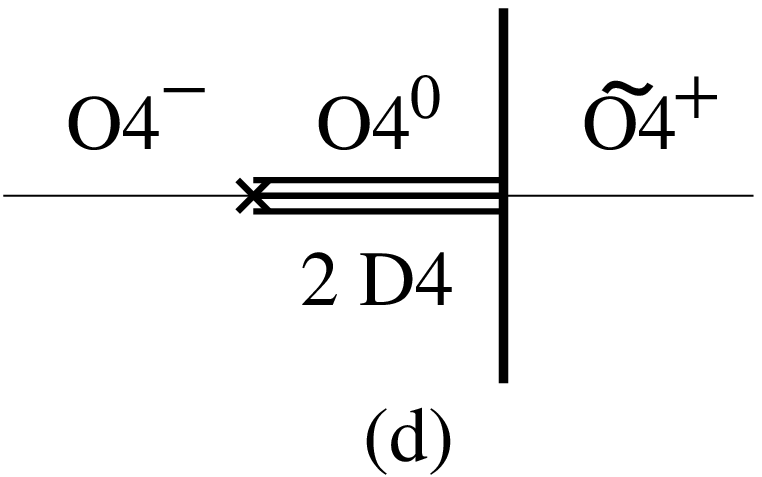}
\end{center}
\caption{S-configurations.}
\label{SR}
\end{figure}
Using a similar argument we can show that
the configurations depicted in Figure \ref{SR}
are not supersymmetric.
Supersymmetry is
also broken if there are larger number of D4-branes in the middle
interval.
Supersymmetry is preserved
if the number of D4-branes in the middle is smaller (and even);
there are six such configurations --- all of them
appear in Figure \ref{Crefig}.

\noindent
(a)~The $x^6>0$ part of the configuration is obtained from the
$\Z_2$ orbifold of (\ref{sco}) with $m=4$ by taking the limit
$\zeta\to 0$ (\ref{2ndlim}),
where the $\Z_2$ acts as (\ref{1stZ2}) fixing the
$x$-axis. If we continue to $x^6<0$, we see that the fivebrane wraps
three-times on the $y$-axis.
In the Type IIA limit (\ref{2ndlim}),
we see that we automatically have two D4-branes on
top of the O4${}^+$-plane on the left of the D6-brane.

\noindent
The proof showing that (b) and (c) are non-supersymmetric
is similar and is not presented here.

\noindent
(d)~
The $x^6>0$ part of the configuration (d)
can be obtained from the $\Z_2$ orbifold
of the configuration with the fivebrane wrapped on
$y=\zeta$
plus another fivebrane wrapped on the $x$-axis,
where the $\Z_2$ acts as (\ref{2ndZ2})
fixing the $y$-axis.
The Type IIA configuration arizes by taking the limit $\zeta\to 0$
(\ref{2ndlim}).
If we look at $x\sim 0$, we see that
the fivebrane wrapping the $x$-axis
(defined as $y=0$) and
the fivebrane wrapped on $y=\zeta$ intersects with the $\Z_2$
fixed plane --- $y$-axis --- at two different points;
$y=\zeta$ and $y=0$.
This is a {\it t}-configuration and is inconsistent by itself.
One way to make it consistent is to deform the two components
so that they intersect at the same point.
If they were deformed, they (or at least one of them)
would no longer be holomorphicaly embedded, and the configuration would
not be supersymmetric.
Another way to make it consistent is to wrap another fivebrane twice
on the $y$-axis. Then, we do not have to deform the original
components and the supersymmetry is preserved. However, 
in the Type IIA limit, there are two D4-branes on top
of O4${}^-$-plane on the left of the D6-brane.

\begin{figure}[htb]
\begin{center}
\epsfxsize=5in\leavevmode\epsfbox{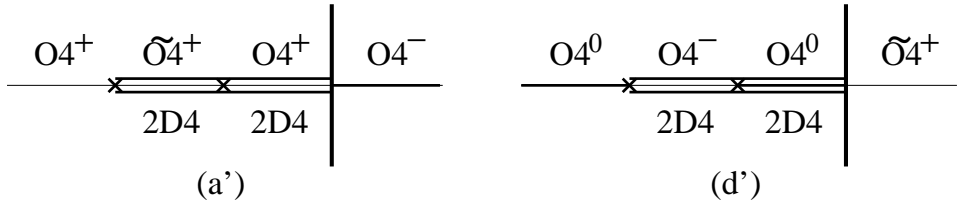}
\end{center}
\caption{Supersymmetric Configurations}
\label{Pres}
\end{figure}

The argument showing that (a) and (d) break supersymmetry implies
also that the configurations depicted in Figure \ref{Pres} preserve
supersymmetry.
We can also see that the configuration (a')
can be obtained by starting from
the bottom-right configuration in Figure \ref{Crefig}
with another D6-brane on the right
and using the brane creation rule.
Similarly, the configuration (d')
can be obtained from the middle-right
configuration in Figure \ref{Crefig} with another
D6-brane on the left, using the brane creation rule,
and exchanging left and right.

\subsection{Type IIA Configuration for $N=2$ SQCD with
$Sp/SO$ Gauge Group}

In the previous subsections, we have derived
the brane creation rule and s-rule in the presence of orientifold
four-planes of various type.
It is an illustrative exercise to see how it
works when we count the dimension of the
Higgs branch of $N=2$ theories using Type IIA
brane configurations.
Since we have already counted it correctly 
using the \MT configuration, and the Type IIA counting is actually
merely a translation of it, we do not present it here.
We only give the Type IIA configuration, leaving the counting as an
exercise for the readers.

\begin{figure}[htb]
\begin{center}
\epsfxsize=5.5in\leavevmode\epsfbox{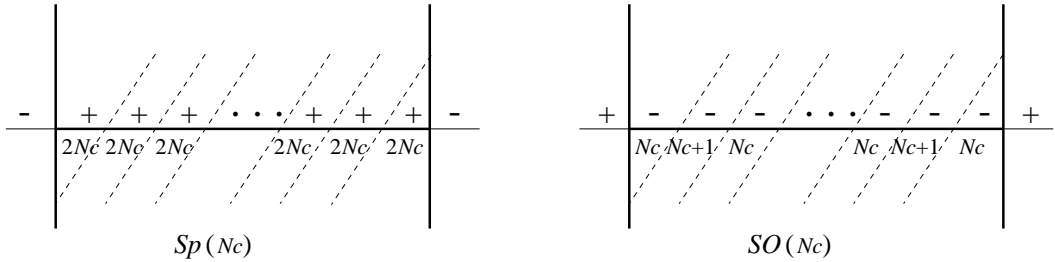}
\end{center}
\caption{Type IIA Configuration for $Sp/SO$ $N=2$ SQCD} 
\label{IIAfig}
\end{figure}

The Type IIA configuration
can be obtained by starting from
the na\"\i ve configuration for the theory with massive quarks,
and using the rule for D6-brane separation as depicted in
Figure \ref{A}.
The same configuration arizes
from the \MT fivebrane configuration given in sections 4 and 5
by taking a suitable limit as in \cite{H}.
The result is depicted in Figure \ref{IIAfig}.
There are two parallel NS5-branes (thick vertical lines)
and $2N_f$ D6-branes (dashed lines) and several D4-branes (thick
horizontal line).
``$+$'' and ``$-$'' written above the horizontal line stand for
$Sp$-type O4-plane and $SO$-type O4-plane
 respectively. The numbers below the line
are the number of D4-branes in the interval.
Note that, in $SO(N_c)$ case, $N_c$ and $N_c+1$ appears alternately.
For even $N_c$, the segment with label $(N_c,-)$ should
be understood as O4${}^-$-plane plus $N_c/2$ pairs of D4-branes while
the $(N_c+1,-)$ piece should
be considered as O4${}^0$-plane plus $N_c/2$ pairs of D4-branes.
The outside O4-planes of $Sp$-type are O4${}^+$ in this case.
For odd $N_c$, the segment with label $(N_c,-)$ should
be understood as O4${}^0$-plane plus $(N_c-1)/2$ pairs of D4-branes while
the $(N_c+1,-)$ segment should
be considered as O4${}^-$-plane plus $(N_c+1)/2$ pairs of D4-branes.
The outside O4-planes of $Sp$-type are $\tilO$ in this case.
In $Sp(N_c)$ case, O4${}^+$ and $\tilO$ appears alternately
as we cross the D6-branes, and in total there are $N_f+1$ O4${}^+$
and $N_f$ $\tilO$.

\bigskip
\bigskip
\noindent{\bf Note Added}

After completing this work, we noticed a paper \cite{Wit}
which contains results substantially related
to the present paper;
a classification of Type IIB orientifold three-plane
is given and intersection of orientifold planes
and various branes is studied. 
The classification says that there are four kind of O3-planes which
have trivial or non-trivial discrete torsion
associated with the field strength of the RR two-form potential
---
$SO({\rm even})$-type with trivial RR torsion,
$SO({\rm odd})$-type with non-trivial RR torsion,
two $Sp$-types with trivial and non-trivial RR torsion.
This pattern looks similar to what we have found in this paper
for O4-planes where the discrete torsion for RR two-form
is replaced by RR $U(1)$ Wilson line.

Whether an orientifold is of $SO$-type or $Sp$-type
can be distinguished by the sign of the
fundamental string $\RP^2$ diagram.
One important thing pointed out in \cite{Wit} is that
this sign reflects the topology of the $B$ field (NS-NS two-form)
which can be characterized by the flux of the
field strength $H=dB$.
In an appendix given below, we show that the \MT realization of
the four O4-planes, as proposed in the present paper,
yield the correct flux of the $H$ field, using the relation
of $B$ field in Type IIA string theory
and the three-form potential $C$ of \MT.

\bigskip
\medskip
\noindent
{\bf Acknowledgement}

I would like to thank O. Bergman,
T. Eguchi, A. Giveon, J. Harvey, S. Sethi, Z. Yin and especially
J. de Boer and C. Vafa for discussions.
I thank D. Freed, A. Givental, R. Kirby and especially
L. Taylor for help and instruction in some topological matters.
I also thank E. Witten for drawing my attention to \cite{Wflux}.
I wish to thank Institute for Theoretical Physics at Santa Barbara,
Physics Department of Harvard University
and the Institute for Advanced Study, for 
hospitality.

This research is supported in part by
NSF grant PHY-95-14797
and
DOE grant DE-AC03-76SF00098,
and also by NSF grant PHY94-07194.

\bigskip
\medskip
\appendix{The Flux of the $H$-Field}

In \cite{Wit}, it was explicitly stated that the sign of the
fundamental string $\RP^2$ diagram surrounding an orientifold
plane (which distinguishes whether the plane is of $SO$ or $Sp$ type)
reflects the topology of the $B$ field which
can be measured by the flux of the $H$ field.
In this appendix, we show that the O4${}^0$ and $\tilO$ constructed in
section 6 have the correct $H$-flux.  For the cases of
O4${}^-$ and O4${}^+$, we uncover a subtlety and show that
one must choose a torsion component of the $G$-flux to specify an
\MT vacuum. We can compute this torsion component when the
O4-plane is intersecting with D6-brane and show
that O4${}^-$ and O4${}^+$ also have the correct
$H$-flux. The result implies that the freezing of
the fivebranes at the $\Z_2$ fixed plane, which is required 
to realize O4${}^+$ plane, is implemented by
the non-vanishing torsion component.

The $B$ field can be obtained from
the three-form potential $C$ of \MT
via integration along the circle in the 
eleventh direction. In other words,
\beq
\int\limits_{S^1} {G\over 2\pi}={H\over 2\pi}.
\label{intfibre}
\eeq
Since the $\Z_2$ action of the double
cover flips the sign of $G$ and preserves
 the orientation of the eleventh circle $S^1$,
the $\Z_2$ action in ten dimensions flips the sign of the $H$ field.
Therefore, $H$ does not define an ordinary three-form on the
quotient but a three-form with values in the orientation bundle.
The topology of the $B$ field is measured by the
cohomology class $[H/2\pi]$ with values in the twisted
integer coefficient $\Zo$ where the twisting is determined by
the orientation bundle. In the present case, the ten-dimensional
space-time (with the orientifold plane deleted to avoid singularity)
is homotopically equivalent to $\RP^4$, and the relevent cohomology
group is
\beq
H^3(\RP^4,\Zo)\cong \Z_2.
\label{RP43}
\eeq
The orientifold plane is of $SO$-type if $[H/2\pi]$ is zero and it is
of $Sp$-type if $[H/2\pi]$ is non-zero.

The eleven-dimensional space-time,
when the locus corresponding to the O4-plane is deleted,
is homotopically equivalent to either
$\RP^4\times S^1$ (for O4${}^-$ or O4${}^+$)
or $(S^4\times S^1)/\Z_2$ (for O4${}^0$ or $\tilO$)
where in the latter case $\Z_2$ maps a point of $S^4$ to its
anti-podal point and acts on the $S^1$ by $\pi$-rotation.
These spaces are fibred over $\RP^4$ in an obvious way
and the integration of $[G/2\pi]$ along the fibre is equal to
$[H/2\pi]$ as in (\ref{intfibre}).
(There is a subtlety
when $w_4$ of eleven-dimensional space-time does not vanish
as we will mention shortly.)

\subsection{The flux of ${\rm O4}^0$ and $\tilO$}

Let us first consider the case of $(S^4\times S^1)/\Z_2$.
The four-th cohomology group of this space
with integer coefficient twisted by the orientation bundle is
\beq
H^4((S^4\times S^1)/\Z_2,\Zo)\cong \Z.
\label{S4S1}
\eeq
Under integration along the fibre $S^1$, it is mapped to
$H^3(\RP^4,\Zo)\cong \Z_2$ as mod 2 reduction,
i.e. even elements are mapped to zero
and odd elements are mapped to the non-zero element.
The submanifold $S$ encountered in section 6.2. (in a sense,
the $S^4$ at a point in $S^1/\Z_2$)
determines a generator of the homology group
$H_4((S^4\times S^1)/\Z_2,\Zo)$ and
integration over this twisted cycle maps the group (\ref{S4S1})
isomorphically onto $\Z$.
Now, in the case of O4${}^0$, since there is no flux of $G$ along
$S$, the cohomology class $[G/2\pi]$ vanishes.
Then, the integration along $S^1$ is of course zero and therefore
$[H/2\pi]=0$. Thus, O4${}^0$ is of $SO$-type.
In the case of $\tilO$, the flux of $G/2\pi$ along $S$ is 1 since
the flux is two in the double cover as we have seen in section 6.2.
Therefore, $[G/2\pi]$ is a generator of (\ref{S4S1}) and integrates
over $S^1$ to the non-zero element of 
$H^3(\RP^4,\Zo)\cong \Z_2$, $[H/2\pi]\ne 0$.
Thus, $\tilO$ is indeed of $Sp$-type.
What has been said is not modified in either case
when even number of fivebranes are
added (in the double cover) since $[G/2\pi]$ gets shifted only
by an even element of (\ref{S4S1}) which has no effect after the
integration along $S^1$.

\subsection{The Flux of ${\rm O4}^-$ and ${\rm O4}^+$}

Let us next consider the case of $\RP^4\times S^1$.
The four-th cohomology group with coefficient twisted by the
orientation bundle is
\beq
H^4(\RP^4\times S^1,\Zo)\cong \Z\oplus \Z_2.
\label{RP4S1}
\eeq
For a suitable decomposition of (\ref{RP4S1}),
the integration along the fibre $S^1$ maps
the $\Z$ factor of (\ref{RP4S1})
to zero and the $\Z_2$ factor isomorphically to
$H^3(\RP^4,\Zo)\cong \Z_2$.
Integration over $\RP^4$ at a point of $S^1$
(which is a twisted four-cycle) maps the $\Z$ factor isomorphically
to $\Z$ and the $\Z_2$ factor to zero.

\subsection*{\sl A Subtle Problem and A Proposal of Resolution}

When we try to use the formula $[H/2\pi]=\int_{S^1}[G/2\pi]$
to measure the $H$-flux of O4${}^-$ and O4${}^+$, we
run into a problem that $G/2\pi$ does not define
a cohomology class in $H^4(\RP^4\times S^1,\Zo)$;
because of
the flux quantization condition (\ref{Flux}) and the fact that
$w_4\ne 0$, $G/\pi$ defines an odd element of
$H^4(\RP^4\times S^1,\Zo)$ which is not divisible by two.
However, if there is a class $c$ in $H^4(\RP^4\times S^1,\Zo)$
which reduces modulo 2 to $w_4$, then by the flux quantization
condition $[G/\pi]-c$ is an even
element of $H^4(\RP^4\times S^1,\Zo)$ which is divisible by two.
Such a class $c$ must satisfy some kind of locality
and desirably be some characteristic class of the (s)pin manifold.
Generically, there is no such characteristic class.\footnote{
Namely, $H^4(BPin(\infty),\Z^{w_1})=0$ where $\Z^{w_1}$ is the twisted
integer coefficient
where the twisting is determined by the canonical unorientable
line bundle.
This is in contrast with the case of
the untwisted coefficient
where there is a ``spin Pontryagin class''
$Q_1\in H^4(B(S)pin(\infty),\Z)$ (with $2Q_1=p_1$) which reduces modulo 2
to $w_4$.}
However, when the structure group of the tangent bundle
is reducible to $O(4)$, there is such a characteristic class
$c$; it is the twisted Euler class $c=\chi$.
Usually Euler class is defined for oriented vector bundle,
as the local expression (in the case of rank 4)
$\chi={1\over 32\pi^2}\epsilon^{ijkl}R_{ij}\wedge R_{kl}$
requires the choice of ``epsilon tensor'' which is globally
defined only when the bundle is orientable. When the vector bundle is
not orientable, the same expression makes sense not as an ordinary
differential form but as a differential form with values
in the orientation bundle. With more care, one can see that it
actually defines a cohomology class with the twisted integer coefficient
$\Zo$ and reduces modulo 2 to $w_4$ as the ordinary Euler class
of oriented bundles.\footnote{
In fact,
$H^4(BPin(n),\Z^{w_1})$ is zero for $n>4$
and is $\Z$ generated by $\chi$ for $n=4$.}
Now, since the structure group of $\RP^4\times S^1$
(or of $\R^5\times (\R^5-0)/\Z_2\times S^1$)
is reducible to $O(4)$, $[G/\pi]-\chi$ is an even element of
$H^4(\RP^4\times S^1,\Zo)$ and is divisible by two.
So, we propose that an \MT vacuum is specified by the choice
of a class
$[\tilde{G}/2\pi]\in H^4(\RP^4\times S^1,\Zo)$ such that
\beq
[G/\pi]=\chi+2[\tilde{G}/2\pi],
\eeq
and that the $H$-flux is obtained by
$[H/2\pi]=\int_{S^1}[\tilde{G}/2\pi]$.

What is $[\tilde{G}/2\pi]$ for O4${}^-$ and O4${}^+$?
The non-torsion part of it
can be determined by the knowledge of $[G/\pi]$ and $\chi$ but
the relevant information for computing $[H/2\pi]$
is in the torsion part. The torsion part is not ``visible''
via $[G/\pi]$ and is an extra information which we must specify to
determine a vacuum.
We recall here that we have not understood
the mechanism of freezing the two fivebranes at the $\Z_2$-fixed plane
to realize the infinite O4${}^+$-plane, and we have not said
what actually distinguishes it from the \MT realization of
the O4${}^-$-plane with two movable D4-branes on top of it.
A most probable solution to the latter problem
would be that they are distinguished by the torsion compoenent of 
$[\tilde{G}/2\pi]$.

\subsection*{\sl The Measurement}

Actually, we can compute the torsion component of $[\tilde{G}/2\pi]$
when the O4-plane in question
ends on a D6-brane and continues to other kind of O4-plane.
Consider the configuration considered in section 7.1. which realize
the O4${}^-$-D6-O4${}^0$ or O4${}^+$-D6-$\tilO$ systems
(note that this is one way to realize O4${}^+$-plane
where we can explain the freezing of the fivebrane
at the $\Z_2$ fixed plane).
The space-time is $\R^4$ times a seven-manifold which is
the Taub-NUT space $yx=v$ times $\R^3=\{x^{7,8,9}\}$ divided by
the $\Z_2$ action $y\to y,x\to -x,v\to -v$
and $x^{7,8,9}\to -x^{7,8,9}$.
In the far left, $|y|$ large, it looks as
$\R^5\times \R^5/\Z_2\times S^1$ (and corresponds to O4${}^-$ or
O4${}^+$) whereas in the far right, $|x|$ large,
it looks as $\R^5\times (\R^5\times S^1)/\Z_2$ (and corresponds to
O4${}^0$ ot $\tilO$).
We focus on the non-trivial seven-manifold piece and
delete the locus $\{v=0,
x^{7,8,9}=0\}$ corresponding to the O4-planes. 
There are three vector fields $\partial/\partial y, \partial/\partial
\bar y$ and $x\partial/\partial x+\bar x\partial/\partial \bar x
+x^7\partial/\partial x^7+x^8\partial/\partial x^8
+x^9\partial/\partial x^9$ which
are well-defined in the $\Z_2$ quotient
and everywhere linearly independent.
Therefore the structure group of the tangent bundle is
reducible to $O(4)$. Thus, one can define the
twisted Euler class $\chi$ and we must choose $[\tilde{G}/2\pi]$
to specify a background.
For a large constant $\alpha$,
the slice with $|y|=\alpha$ is homotopically equivalent
to $\RP^4\times S^1$
while the slice with $|x|=\alpha$ is homotopic to $(S^4\times S^1)/\Z_2$.
The class $[\tilde{G}/2\pi]$, when restricted to these
slices, defines cohomology classes with the coefficient $\Zo$
of these spaces.
The torsion component of $[\tilde{G}/2\pi]$ for the semi-infinite
O4${}^-$ or O4${}^+$ can be measured by evaluating it over
the cycle $\RP^3\times S^1$ in the slice $|y|=\alpha$.
(This cycle is an untwisted cycle but the
integral of $[\tilde{G}/2\pi]$ on it is well-defiend
as a mod 2 integer.)
There is actually a homotopy connecting this cycle and
the cycle $(S^3\times S^1)/\Z_2$
in the slice $|x|=\alpha$,\footnote{For example,
$|y|^2+|x|^2+|x^{8,9}|^2=\alpha^2+\beta^2$ and $x^7=0$,
with $|y|, |x| \leq \alpha$ for some $0<\beta<\alpha$.}
and therefore the flux in question is measured by integrating
$[\tilde{G}/2\pi]$ over $(S^3\times S^1)/\Z_2$.
Since $w_4$ of $(\R^5\times S^1)/\Z_2$ is zero, $\chi$ is even
on $(S^4\times S^1)/\Z_2$ and thus we may replace
$[\tilde{G}/2\pi]$ by $[G/2\pi]$ on this side.
Now, let us consider the O4${}^-$-D6-O4${}^0$
system. Since $[G/2\pi]=0$ for O4${}^0$, using the homotopy,
we see that the integral of $[\tilde{G}/2\pi]$
over $\RP^3\times S^1$ is zero. Namely, $[\tilde{G}/2\pi]$
for O4${}^-$
has no torsion component in the decomposition (\ref{RP4S1}).
Next, we consider the O4${}^+$-D6-$\tilO$ system.
Since $\int_{S^1}[G/2\pi]$ is non-zero for $\tilO$
as we have seen above, the integral over $(S^3\times S^1)/\Z_2$
is non-zero, and therefore the integral of
$[\tilde{G}/2\pi]$ over $\RP^3\times S^1$ is non-zero.
Namely, $[\tilde{G}/2\pi]$ for O4${}^+$
has the non-zero torsion component in
(\ref{RP4S1}).

To summarize, the $\Z_2$ component of $[\tilde{G}/2\pi]$
in the decomposition (\ref{RP4S1}) is zero for the semi-infinite
O4${}^-$
but non-zero for the semi-infinite O4${}^+$.
In either case,
this is also true when even number of D4-branes are added.
Thus, O4${}^-$ is of $SO$-type while O4${}^+$ is of $Sp$-type
as far as they are obtained by sending the D6-brane to the right
infinity.

\subsection*{\sl The Torsion and the Freezing of Fivebranes}

One of the interesting fact uncovered through this 
discussion is that the freezing of a pair of fivebranes at 
the $\Z_2$ fixed plane (with topology $\R^5\times S^1$)
is correlated with the $\Z_2$ torsion component of
$[\tilde{G}/2\pi]$ in the decomposition (\ref{RP4S1}).
We still do not understand the freezing mechanism when the
O4${}^+$-plane is infinite. However, in the case considered above
where O4${}^+$ is semi-infinite,
the $\Z_2$ fixed plane is intersecting with a
single fivebrane, and
the question of freezing is traced back to the consistency
condition studied in section 2 starting with the flux quantization
condition (\ref{Flux}). In this way we can explain why the freezing
is correlated with the non-vanishing $\Z_2$ torsion of
$[\tilde{G}/2\pi]$.

As mentioned in section 6, such a mysterious freezing has been
observed for an O6-plane \cite{LL} as frozen $D_4$
singularity and for an
O7-plane \cite{Wtor} as frozen $D_8$ singularity.
In the latter case, correlation of the
freezing and the flux of a $B$-field
has been pointed out. A generalization of this has been discussed
in \cite{BPS,BKMT}.

\end{document}